%% file: opinion.tex
\documentclass[final,11pt,5p,twocolumn,sort&compress]{elsarticle}

\usepackage{amsfonts,amsmath,amssymb}
\usepackage{bbm}
\usepackage{epstopdf}
\usepackage{caption,subcaption}

\journal{Annual Reviews in Control}
\usepackage{lineno,hyperref}

\def\diag{\mathop{\rm diag}\nolimits}
\def\E{\mathbb{E}}

\def\la{\lambda}
\def\La{\Lambda}
\def\vp{\varphi}
\def\ve{\varepsilon}

\def\r{\mathbb R}
\def\cc{\mathbb C}

\def\C{\mathcal C}

\def\g{\mathcal G}
\def\P{\mathbb{P}}
\def\E{\mathbb{E}}
\def\I{\mathbb{I}}

\newcommand{\dfb}{\stackrel{\Delta}{=}}

\def\be{\begin{equation}}
\def\ee{\end{equation}}
\def\ben{\begin{equation*}}
\def\een{\end{equation*}}

\def\sgn{\mathop{sgn}\nolimits}

\newtheorem{thm}{Theorem}
\newtheorem{lem}[thm]{Lemma}
\newtheorem{prop}[thm]{Proposition}
\newtheorem{rem}[thm]{Remark}
\newtheorem{cor}[thm]{Corollary}
\newdefinition{defn}{Definition}

\def\ones{\mathbbm{1}}
\def\re{\mathop{\rm Re}\nolimits}
\def\imath{\mathrm{\bf i}}

\begin{document}

\begin{frontmatter}

\title{A Tutorial on Modeling and Analysis of Dynamic Social Networks. Part II.\tnoteref{t}}
\tnotetext[t]{The paper is supported by Russian Science Foundation (RSF) grant 14-29-00142, hosted by IPME RAS.}

\author[TUD,IPME,ITMO]{Anton V. Proskurnikov\corref{mycorrespondingauthor}}
\ead{anton.p.1982@ieee.org}

\author[CNR]{Roberto Tempo}
\cortext[mycorrespondingauthor]{Corresponding author}

\address[TUD]{Delft Center for Systems and Control, Delft University of Technology, Delft, The Netherlands}
\address[IPME]{Institute for Problems of Mechanical Engineering of the Russian Academy of Sciences, St. Petersburg, Russia}
\address[ITMO]{ITMO University, St. Petersburg, Russia}
\address[CNR]{CNR-IEIIT, Politecnico di Torino, Torino, Italy}

\begin{abstract}
Recent years have witnessed a significant trend towards filling the gap between Social Network Analysis (SNA) and control theory. This trend was enabled by the introduction of new mathematical models describing dynamics of social groups, the development of algorithms and software for data analysis and the tremendous progress in understanding complex networks and multi-agent systems (MAS) dynamics. The aim of this tutorial is to highlight a novel chapter of control theory, dealing with dynamic models of social networks and processes over them, to the attention of the broad research community. In its first part~\citep{ProTempo:2017-1}, we have considered the most classical models of social dynamics, which have anticipated and to a great extent inspired the recent extensive studies on MAS and complex networks.
This paper is the second part of the tutorial, and it is focused on more recent models of social processes that have been developed concurrently with MAS theory. Future perspectives of control in social and techno-social systems are also discussed.
\end{abstract}

\begin{keyword}
Social network, opinion dynamics, multi-agent systems, distributed algorithms.
\end{keyword}

\end{frontmatter}


\input 1intro
\input 2prelim
\input 3krause

\input 4gossip
\input 5negative
\input 6conclus

\section*{References}

\bibliographystyle{elsarticle-num}
\bibliography{social,networks,consensus}

\end{document}

%% file: 1intro.tex
\section{Introduction}

Originating from the early studies on \emph{sociometry}~\citep{Moreno1934,Moreno1951}, Social Network Analysis (SNA) has quickly grown into an interdisciplinary science~\citep{WassermanFaustBook,Scott_Handbook2000,Freeman2004,HandBookSNA2011} that has found applications in political sciences~\citep{KnokeBook:1990,Lazer:2011}, medicine~\citep{OMalleyMarsden:2008}, economics~\citep{EasleyKleinberg,JacksonBook2008}, crime prevention and security~\citep{Defense-2014,Criminal-2015}. The recent breakthroughs in algorithms and software for big data analysis have made SNA an efficient tool to study online social networks and media~\cite{ArnaboldiBook_OSN,KazienkoChawla-2015} with millions of users. The development of SNA has inspired many important concepts of modern \emph{network science}~\citep{Strogatz:2001,NewmanReview:2003,NewmanBarabasiBook,VanMieghemBook} such as cliques and communities, centrality measures, resilience, graph's density and clustering coefficient.

Employing many mathematical and algorithmic tools, SNA has however benefited little from the recent progress in systems and control~\citep{MurrayReport,SamadReport,IFACAgenda2017}. The realm of social sciences has remained almost untouched by control theory, despite the long-term studies on
social group dynamics~\citep{Lewin:1947,SorokinBook1947,DianiAdamBook} and~``sociocybernetics''~\citep{WienerBook-Socio,Geyer:1995,GeyerBook_2001,Bailey:2006}.
This gap between SNA and control can be explained, to a great extent, by the lack of dynamic \emph{models} of social processes and mathematical armamentarium for their analysis.
Focusing on topological properties of networks, SNA and network science have paid much less attention to dynamics over them, except for some
special processes such as e.g. random walks, branching and queueing processes, percolation and contagion dynamics~\citep{NewmanBarabasiBook,VanMieghemBook}.

The recent years have witnessed an important tendency towards filling the gap between SNA and control theory, enabled by the rapid progress in multi-agent systems and dynamic networks.
The emerging branch of control theory, studying social processes, is very young and even has no name yet. However, the interest of sociologists to this new area and
understanding that ``coordination and control of social systems is the foundational problem of sociology''~\cite{Friedkin:2015} leaves no doubt that it should become a key instrument to examine social networks and dynamics over them. Without aiming to provide a exhaustive survey of ``social control theory'' at its dawn, this tutorial focuses on the most ``mature'' results, primarily dealing with mechanisms of \emph{opinion formation}~\citep{Holyst:2001,Castellano:2009,AcemogluDahleh:2011,XiaWangXuan:2011,Friedkin:2015,DongZhanKou:18-survey}.

In the first part of this tutorial~\citep{ProTempo:2017-1}, the most classical models of opinion formation have been discussed that
have anticipated and inspired the ``boom'' in multi-agent and networked control, witnessed by the past decades.
This paper is the second part of the tutorial and deals with more recent dynamic models, taking into account effects of time-varying graphs, homophily, negative influence, asynchronous interactions and quantization. The theory of such models and multi-agent control have been developed concurrently, inspiring and reinforcing each other.

Whereas analysis of the classical models addressed in~\citep{ProTempo:2017-1} is mainly based on linear algebra and matrix analysis, the models discussed in this part of the tutorial
require more sophisticated and diverse mathematical tools. The page limit makes it impossible to include the detailed proofs of all results discussed in this part of the tutorial;
for many of them, we have to omit the proofs or provide only their brief sketches.

The paper is organized as follows. Section~\ref{sec.prelim} introduces preliminary concepts and some notation used throughout the paper.
Section~\ref{sec.time-var} considers basic results, concerned with properties of the non-stationary French-DeGroot and Abelson models.
In Section~\ref{sec.krause} we consider \emph{bounded confidence} models, where the interaction graph is \emph{opinion-dependent}.
Section~\ref{sec.gossip} is devoted to dynamic models based on asynchronous \emph{gossiping} interactions. Section~\ref{sec.negative} introduces some models, exploiting the idea of \emph{negative influence}. Section~\ref{sec.conclus} concludes the tutorial. 

%% file: 2prelim.tex
\section{Preliminaries and notation}\label{sec.prelim}

In this section we introduce some notation; basic concepts regarding opinion formation modeling
are also recollected for the reader's convenience.

\subsection{Notation}

We use $m:n$ to denote the set $\{m,m+1,\ldots,n\}$ (here $m,n$ are integer and $m\le n$). Given a vector $x\in\r^n$, $|x|$ stands for its Euclidean norm $|x|=\sqrt{x^{\top}x}$.

Each non-negative matrix $A=(a_{ij})_{i,j\in V}$ corresponds to the weighted graph $\g[A]=(V,E[A],A)$, whose arcs represent positive entries
of $A$: $a_{ij}>0$ if and only if $(j,i)\in E(A)$. Being untypical for graph theory (where the entry $a_{ij}>0$ is encoded by the arc $(i,j)$), this notation is convenient
in social dynamics modeling~\cite{ProTempo:2017-1} and multi-agent control~\cite{RenBeardBook,RenCaoBook}.

Dealing with algorithms' complexity, we use standard Landau-Knuth notation~\cite{Knuth_Notation}. Given two positive functions $f(n),g(n)$ of the natural argument $n$,
$g(n)=O(f(n))$ stands for the estimate $|g(n)|\le C|f(n)|$, where $C$ is some constant, and $f(n)=\Omega(g(n))$ means that $\varliminf\limits_{n\to\infty} f(n)/g(n)>0$ (i.e. $f(n_k)\ge c_0g(n_k)$ for a constant $c_0>0$ and a sequence $n_k\to\infty$.

\subsection{Agent-based modeling of opinion evolution}


From the sociological viewpoint~\citep{Friedkin:2015}, an individual's opinion stands for his/her \emph{cognitive orientation} towards some object (e.g. issue, event, action or another individual). Mathematically, opinions are scalar or vector quantities of interest, associated with social actors. Depending on the specific model, opinions may represent signed attitudes~\citep{Abelson:1964,HunterDanesCohenBook_1984,Kaplowitz:1992}, subjective certainties of belief~\cite{Halpern:1991,FriedkinPro2016} or probabilities~\citep{DeGroot,Scaglione:2013}. In this tutorial, we deal with models where opinions can attain a \emph{continuum} of values and are represented by real numbers or vectors. Dynamics of real-valued opinions obey ordinary differential or recurrent equation and are much better studied by the systems and control community than the evolution of discrete (finite-valued) opinions.
For this reason, many important models with finite-valued opinions~\citep{Clifford:1973,Holley:1975,Granovetter:1978,CoxGriffeath:1983,Axelrod:1997,Sznajd:2000,Yildiz:2013,DongChen:16,TupikinaArxiv:17}
are beyond the scope of this tutorial.

Historically, the first approach to social dynamics modeling originates from mathematical biology~\citep{Rashevsky:1935,Rashevsky:1939,Rashevsky1938Bio,Rashevsky1947}, portraying social behaviors
as interactions of multiple ``species'' or \emph{compartments}~\cite{Jacquez:1985}. Dealing with a social group, a compartment is a subgroup whose members are featured by some behavior
or hold the same position on some issue. Interacting as indecomposable entities, compartments can grow or decline. The models describing these processes are called \emph{compartmental} and broadly used in mathematical biology and evolutionary game theory~\cite{Jacquez:1985,Epidemiology_Book,SmithBook1982_EvolGames}, as exemplified by the SIR/SIS models of epidemic spread and the Lotka-Volterra predator-prey dynamics. Compartmental models describe how the \emph{distribution} of individuals between the compartments evolves, paying no attention to behaviors of
specific social actors. This \emph{statistical} approach is typical for \emph{sociodynamics}~\citep{Weidlich:1971,Weidlich:2005,Helbing:2010},
representing the state of a society by a point in some \emph{configuration space} and has lead to statistical model of opinion formation, describing how the distribution of opinions evolves over time. Similar in spirit to models arising in continuum mechanics, such models are often referred to as \emph{Eulerian}~\cite{ComoFagnani:2011,CanutoFagnani:2012,MirtaJiaBullo:2014} or \emph{continuum}~\citep{Blondel:2009,WedinHegarty:2015,HendrickxOlshevsky:16} opinion dynamics.

In this tutorial, we focus on \emph{agent-based} models of opinion formation, describing how the opinion of each individual social actor, or \emph{agent}, evolves under the influence of the remaining individuals. The collective behavior of a social group is constituted by the numerous individual behaviors. Such ``bottom-up'' models of social dynamics, called also aggregative~\citep{Abelson:1967}, are similar in spirit to agent-based models of self-organization in complex physical and biological systems~\cite{Reynolds,Vicsek,StrogatzSync}.
Unlike statistical models, adequate for very large social groups, agent-based models can describe both small-size and large-scale communities.
Throughout this paper, we deal with a closed community of $n\ge 2$ agents, indexed $1$ through $n$.

\subsection{Models of consensus and Abelson's puzzle}

As have been discussed in the first part of this tutorial~\citep{ProTempo:2017-1}, the first agent-based model of opinion formation was introduced by French~\citep{French:1956} and later
studied and extended by Harari~\citep{Harary:1959,HararyBook:1965} and DeGroot~\cite{DeGroot}. The French-DeGroot model describes the discrete-time evolution of the agents opinions $x_1,\ldots,x_n\in\r$, whose stack vector $x(k)=(x_1(k),\ldots,x_n(k))^{\top}\in\r^n$ at step $k=0,1,\ldots$ obeys the averaging dynamics
\be\label{eq.degroot}
x(k+1)=Wx(k),\quad k=0,1,\ldots
\ee
where $W=(w_{ij})$ is a stochastic matrix. The continuous-time counterpart of~\eqref{eq.degroot}, proposed by Abelson~\cite{Abelson:1964}, is the \emph{Laplacian flow} dynamics~\cite{BulloBook-Online}
\be\label{eq.abelson}
\dot x(t)=-L[A]x(t),\quad t\ge 0,
\ee
where $A=(a_{ij})$ is a non-negative matrix of ``contact rates'' and $L[A]$ stands for the Laplacian matrix of the corresponding weighted graph~\cite{ProTempo:2017-1,BulloBook-Online}.
The asymptotic consensus of opinions appears to be the most typical behavior of the systems~\eqref{eq.degroot} and~\eqref{eq.abelson}, the relevant criteria are considered in~\cite{ProTempo:2017-1}.
At the same time, real social groups often fail to reach consensus and exhibit clustering of opinions and other ``irregular'' behaviors. This has lead Abelson to the fundamental problem, called the
\emph{community cleavage} problem~\cite{Friedkin:2015} or Abelson's \emph{diversity puzzle}~\cite{Nakamura:16}:
to find mathematical models, able to explain these disagreement effects. The original formulation of Abelson~\citep{Abelson:1964} was as follows: ``we are naturally lead to inquire what on earth one must assume in order to generate the bimodal outcome of community cleavage studies''.

One reason for community cleavage is the absence of connectivity: consensus of opinions in the models~\eqref{eq.degroot} and~\eqref{eq.abelson} cannot be established when the corresponding interaction graph $\g[W]$ or $\g[A]$ has no directed spanning tree. Although social networks are usually densely connected~\cite{Watts6Degrees}, they may contain
some ``radical'' groups~\cite{HegselmannKrause:2015}, closed to social influence. For instance, consensus is not possible in presence for several \emph{stubborn} individuals (or \emph{zealots})~\cite{Yildiz:2013,Masuda:2015}, whose opinion remains unchanged $x_i\equiv x_i(0)$. Further development of this idea naturally leads~\cite{ProTempo:2017-1}
to the Friedkin-Johnsen theory of social influence networks~\cite{FriedkinJohnsen:1999,FriedkinJohnsenBook} with ``partially stubborn'' agents.

Stubborness is however not the only factor leading to the community cleavage; in this part of the tutorial we consider other models of opinion formation where opinions can both converge to consensus or split into several clusters. Many of these models are based on the ideas, proposed in the seminal Abelson's works~\citep{Abelson:1964,Abelson:1967} and extend the classical models~\eqref{eq.degroot},\eqref{eq.abelson}.

\section{The models by French-DeGroot and Abelson with time-varying interaction graphs}\label{sec.time-var}

Non-stationary counterparts of the models~\eqref{eq.degroot} and~\eqref{eq.abelson} have been thoroughly studied in regard to consensus and synchronization in multi-agent networks.
In this tutorial, only some results are considered that directly related to social dynamics; detailed overview of consensus algorithms can be found e.g. in the recent monographs and surveys~\cite{Murray:07,LiDuanChen:10,RenBeardBook,RenCaoBook,Wu2007,ProCao16-EEEE,ProFradkov:2016}.

\subsection{The time-varying French-DeGroot model}

We start with a time-varying counterpart of~\eqref{eq.degroot}, where $W$ is replaced by a sequence $(W(k))_{k\ge 0}$
\be\label{eq.degroot1}
x(k+1)=W(k)x(k),\quad k=0,1,\ldots
\ee
Obviously, all solutions to~\eqref{eq.degroot1} are bounded and the sequences $\min_i x_i(k)$ and $\max_i x_i(k)$, $k=0,1,\ldots$, are respectively non-decreasing and non-increasing.

As discussed in~\cite{ProTempo:2017-1}, even for the static case $W(k)\equiv W$ the opinions do not always converge. For instance, when the graph $\g[W]$ is \emph{periodic}, the system~\eqref{eq.degroot} has a periodic solution. The convergence problem for time-varying system~\eqref{eq.degroot1} still remains a challenge, and up to now only sufficient convergence conditions have been obtained. One of them is given by the following important result, proved independently in~\cite{Blondel:05,Moro:05,Lorenz:2005}.
\begin{lem}\label{lem.converge1}
Let $\delta>0$ exist such that the sequence of $n\times n$ stochastic matrices $(W(k))_{k\ge 0}$ satisfies at any time $k\ge 0$ the following three conditions:
\begin{enumerate}[(a)]
\item (non-vanishing couplings) $w_{ij}(k)\in\{0\}\cup [\delta,1]$;
\item (self-confidence) $w_{ii}(k)\ge\delta\quad\forall i\in 1:n$;
\item (type-symmetry) $w_{ij}(k)>0\Longleftrightarrow w_{ji}(k)>0$.
\end{enumerate}
Then the limit $\bar x=\lim_{k\to\infty} x(k)$ exists for any $x(0)\in\r^n$, being an equilibrium point: $W(k)\bar x=\bar x$ for sufficiently large $k$.
If agents $i$ and $j$ \emph{interact persistently}
\[
\sum_{k=0}^{\infty} w_{ij}(k)=\infty,
\]
then their limit opinions coincide $\bar x_i=\bar x_j$.
\end{lem}

Introducing the undirected \emph{graph of persistent interactions} $\g_{\infty}=(V,E_{\infty})$, whose nodes stand for the agents and arcs $(i,j)$ connect pairs of persistently interacting agents, the last statement of Lemma~\ref{lem.converge1} can be formulated as follows: \emph{in each connected component of $\g_{\infty}$, the opinions reach consensus}.

We give a sketch of the proof of Lemma~\ref{lem.converge1}, following the ideas from~\cite{Blondel:05} and proposed in~\cite{ProCao2017-3} for more general systems of \emph{recurrent inequalities}.
It suffices to consider the case of \emph{connected} graph $\g_{\infty}$. Indeed, if $(i,j)\not\in E_{\infty}$, then $w_{ij}(k)>0$ only for finite number of instants $k$ thanks to condition (a). In other words, $k_0\ge 0$ exists such that $w_{ij}(k)=0$ for $k>k_0$ unless $i$ and $j$ persistently interact. Renumbering the agents, for $k>k_0$ the matrix $W(k)$ is block diagonal
\[
W(k)=\begin{pmatrix}
W_{11}(k) & \ldots & 0\\
\vdots & \ddots & \vdots\\
 0  & \ldots & W_{rr}(k)
\end{pmatrix},
\]
where the stochastic submatrices $W_{ii}(k)$ correspond to connected components of $\g_{\infty}$. Hence~\eqref{eq.degroot1} for $k>k_0$ is decoupled into $r$ independent systems.

Let $\g_{\infty}$ be connected and $j_1(k),\ldots,j_n(k)$ be the permutation of indices, sorting the opinions $x_1(k),\ldots,x_n(k)$ in the ascending order, that is,
$y_i(k)=x_{j_i(k)}(k)$ satisfy the following inequalities
\[
\min_i x_i(k)=y_1(k)\le y_2(k)\le\ldots\le y_n(k)=\max_i x_i(k).
\]
We are going to prove the following statement: \emph{for any $k\ge 0$ and $i=1,\ldots,n-1$, there exists $k'>k$ (depending on both $k$,$i$), satisfying the inequality
\be\label{eq.aux1}
y_{i+1}(k')\le \delta y_i(k)+(1-\delta) y_n(k),
\ee
where $\delta>0$ is the constant from condition (a).} To prove this, divide the agents into two sets $I=\{j_1(k),\ldots,j_i(k)\}$ and $J=\{j_{i+1}(k),\ldots,j_n(k)\}$.
Since $\g_{\infty}$ is connected, an arc between $I$ and $J$ should exist, and hence there exist $K>k$, such that $w_{qp}(K)\ge\delta$ for some $p\in I,q\in J$.
Let $k_0$ stand for the minimal such $K$. Since $x_s(k)\le y_i(k)$ for any $s\in I$ and the agents from $I$ and $J$ do not interact at times $k,k+1,\ldots,k_0-1$, it can be shown that
$x_r(k_0)\le y_i(k)\,\forall r\in I$. Also $x_r(k_0)\le y_n(k_0)\le y_n(k)\,\forall r\in J$ since $y_n(k)$ is non-increasing in $k$. For $r\in I$, one has
\[
\begin{split}
x_r(k_0+1)\le x_r(k_0)+\overbrace{\sum_{l\ne r}w_{rl}(k_0)}^{1-w_{rr}(k_0)\le 1-\delta}[\underbrace{x_l(k_0)}_{\le y_n(k_0)}-x_r(k_0)]\le \\
\le x_r(k_0)+(1-\delta)[y_n(k_0)-x_r(k_0)]\le \\ \le \delta x_r(k_0)+(1-\delta) y_n(k_0)\le \delta y_i(k)+(1-\delta) y_n(k).
\end{split}
\]
Recalling that $w_{qp}(k_0)\ge \delta$ and $p\in I$, similarly to the previous inequality one obtains
\[
\begin{split}
x_q(k_0+1)\le w_{qp}(k_0)x_p(k_0)+(1-w_{qp}(k_0))y_n(k_0)= \\
= y_n(k_0)-w_{qp}(k_0)[y_n(k_0)-x_p(k_0)]\le \\
\le \delta x_p(k_0)+(1-\delta) y_n(k_0)\le \delta y_i(k)+(1-\delta) y_n(k).
\end{split}
\]
Denoting $k'=k_0+1$, for any index $\rho\in I'=I\cup\{q\}$ one has $x_{\rho}(k')\le \delta y_i(k)+(1-\delta) y_n(k)$. Since $I'$ contains $i+1$ different indices, one arrives at~\eqref{eq.aux1}.
Since $y_n(k)$ is bounded from below and non-increasing, it converges to a limit $y_n(k)\to M_*$ as $k\to\infty$.
Passing to the limit as $k\to\infty$ in~\eqref{eq.aux1}, the corresponding sequence $k'=k'(i,k)$ also tends to $\infty$ and thus
\[
\varliminf_{k\to\infty} y_{i+1}(k)\le\delta \varliminf_{k\to\infty} y_{i+1}(k)+(1-\delta)M_*.
\]
Applying this to $i=n-1$, one has $M_*\le\varliminf_{k\to\infty} y_{n-1}(k)\le \varlimsup_{k\to\infty} y_{n-1}(k)\le\varlimsup_{k\to\infty} y_{n}(k)=M_*$, and therefore $y_{n-1}(k)\xrightarrow[k\to\infty]{} M_*$. Iterating this procedure for $i=n-2,\ldots,1$, one proves that $y_{i}(k)\to M_*$, i.e. consensus of opinions is established.
Obviously, any consensus vector $c\ones_n$ is an equilibrium point, which finishes the proof. $\blacksquare$

The convergence of opinions in~\eqref{eq.degroot1} can be reformulated in terms of matrix products convergence~\cite{Lorenz:2005}.
\begin{cor}\label{cor.product}
Under the assumptions of Lemma~\ref{lem.converge1}, the limit of the matrix products exist
\be\label{eq.product}
\bar W=\lim_{k\to\infty} W(k)\ldots W(1)W(0).
\ee
Renumbering of the agents, $\bar W$ is block-diagonal
\[
\bar W=\begin{pmatrix}
\bar W_{11} & \ldots & 0\\
\vdots & \ddots & \vdots\\
 0  & \ldots & \bar W_{rr}
\end{pmatrix},
\]
where $r$ is the number of connected components in $\g_{\infty}$ and $\bar W_{ii}$ are ``consensus matrices'' $\ones_{n_i}p_i^{\top}$, where $p_i\in\r^{n_i}$ is a non-negative vector\footnote{Similar to the static model~\eqref{eq.degroot}, the elements of $p_i\ge 0$ can be considered as \emph{social powers} of the corresponding agents~\cite{ProTempo:2017-1}.} with $p_i^{\top}\ones_{n_i}=1$.
\end{cor}

It should be noticed that the \emph{consensus} criterion from Lemma~\ref{lem.converge1} can be substantially extended~\cite{ShiJohansson:13-1,XiaShiMengCaoJohansson:2017}. In particular,
are reported in the preprint~\cite{XiaShiMengCaoJohansson:2017}, in the case of connected undirected graph $\g_{\infty}$ the type-symmetry condition (c) can be relaxed to its ``non-instantaneous''
version and (a) can be discarded (assuming (b) to be valid). Many models of opinion dynamics (e.g. time-varying extensions of the FJ model~\cite{ProTempoCao16-2}), however, exhibit disagreement instead of consensus, being uncovered by these strong results.

A natural question arises how to measure the rate of convergence in~\eqref{eq.degroot1} (being, in general, non-exponential). Possible measures for such a rate are the total and the kinetic \emph{$s$-energies}~\cite{Chazelle:11}, defined as
\be\label{eq.s-energy}
\begin{gathered}
E(s)=\sum_{\substack{k=0,1,2,\ldots\\(i,j):w_{ij}(k)\ne 0}}|x_i(k)-x_j(k)|^s\\
K(s)=\sum_{k=0}^{\infty}\sum_{i=1}^n|x_{i}(k+1)-x_i(k)|^s.
\end{gathered}
\ee
It is not obvious that $E(s)$ and $K(s)$ are finite for $s\le 1$, however, under the assumptions of Lemma~\ref{lem.converge1} both series converge\footnote{Choosing the ``agreement parameter'' $\rho=\min(\delta,1/2)$, the conditions (a),(b) entail the inequalities (3) from~\cite{Chazelle:11}.} for all $s>0$.
Some explicit estimates for $E(s)$ and $K(s)$, depending on $s,n,\delta$ and the initial condition $x(0)$, has been derived in~\cite{Chazelle:11}.

In practice, the assumption (c) restricting the interactions among the agents to be bidirectional (or \emph{reciprocal}) often fails.
The dynamics of matrix products without the type-symmetry assumption is a long-standing problem in matrix analysis and non-stationary Markov chain theory~\cite{Hajnal:58,Wolfowitz:1963,Seneta:77}. Some extensions of Corollary~\ref{cor.product} to the matrices without type-symmetry assumption has been reported in~\cite{Lorenz:2006,TsiTsi:13}.
Most of the existing convergence criteria are however confined to the case where the model~\eqref{eq.degroot1} exhibits consensus. The following fundamental property, established in~\cite{Seneta:77},
shows the equivalence between the ``weak'' and ``strong'' definitions of consensus, existing in the literature and corresponding to, respectively, weak and strong \emph{ergodicity}~\cite{Seneta:77} of the backward matrix products $W(k)\ldots W(0)$.
\begin{prop}\label{prop.ergo}
For any sequence $\{W(k)\}$, the following two conditions are equivalent:
\begin{enumerate}
\item for any $x(0)$, opinions asymptotically synchronize so that $\max_{i,j}|x_i(k)-x_j(k)|\xrightarrow[k\to\infty]{}0$;
\item for any $x(0)$, the opinions converge to a common limit $x_i(k)\xrightarrow[k\to\infty]{} x_*$ (which depends on $x(0)$).
\end{enumerate}
\end{prop}

The following lemma gives a widely known \emph{sufficient} condition for consensus~\cite{Blondel:05,Moro:05,RenBeardBook,CaoMorse:08Part1}.
\begin{lem}\label{lem.consensus1}
Suppose that the sequence of $n\times n$ stochastic matrices $(W(k))_{k\ge 0}$ satisfies the conditions (a) and (b) from Lemma~\ref{lem.converge1}. Additionally, let the following
\emph{repeated quasi-strong connectivity} hold: there exists $T>0$ such that the following graphs
\be\label{eq.graph-union1}
\g_{T}(k)=\g[W(k)+\ldots+W(k+T-1)],\quad k\ge 0
\ee
have directed spanning trees (quasi-strongly connected). Then the opinions \emph{exponentially} converge to consensus $x(k)\xrightarrow[k\to\infty]{}c\ones_n$, where $c=c(x(0))\in\r$.
\end{lem}

The repeated quasi-strong connectivity implies that the \emph{union} of each $T$ consecutive graphs is quasi-strongly connected, extending thus the consensus criterion for static French-DeGroot model~\eqref{eq.degroot}~\cite[Corollary~13]{ProTempo:2017-1}.
Lemmas~\ref{lem.converge1} and \ref{lem.consensus1} remain valid in presence of \emph{communication delays}~\cite{Blondel:05,Bliman:06} and can be extended to some nonlinear consensus algorithms~\cite{Moro:05,Bliman:06,Pro13AutDelay}.

\subsection{The time-varying Abelson model}

The convergence criterion, similar to Lemma~\ref{lem.converge1}, holds also for the time-varying counterpart of~\eqref{eq.abelson}
\be\label{eq.abelson1}
\dot x(t)=-L[A(t)]x(t),\quad t\ge 0.
\ee
Here $A(t)=(a_{ij}(t))$ is a non-negative matrix, whose entries are suppose to be locally $L_1$-summable. Unlike the static Abelson model with $A(t)\equiv A$, where opinions always converge~\cite{ProTempo:2017-1}, the convergence of~\eqref{eq.abelson1} is a non-trivial problem. For the case of bidirectional (reciprocal) interactions, however, the following elegant result has been obtained in~\cite{TsiTsi:13,MatvPro:2013}.
\begin{lem}\label{lem.converge2}
Suppose that the gains $a_{ij}(t)$ satisfy the following type-symmetry condition
\be\label{eq.symmetry}
K^{-1}a_{ji}(t)\le a_{ij}(t)\le Ka_{ji}(t)\quad \forall t\ge 0,
\ee
where $K\ge 1$ is a constant. Then the functions $\dot x_j,a_{ij}(x_j-x_i)$ are $L_1$-summable for any $i,j$. In particular,
the limit $\bar x=\lim_{k\to\infty} x(k)$ exists and if agents $i$ and $j$ \emph{interact persistently} in the sense that
\[
\int_{0}^{\infty} a_{ij}(t)dt=\infty,
\]
then their final opinions are coincident $\bar x_i=\bar x_j$.
\end{lem}

The proofs in~\cite{TsiTsi:13,MatvPro:2013} are based on the properties of the ordering permutation $y_i(t)=x_{j_i(t)}(t)$, sorting the opinions in the ascending order, we do not include them
as they require some non-trivial mathematical tools.
In the case where consensus is established (the graph of persistent interactions is connected), an alternative proof has been proposed in~\cite{ShiJohansson:13}.
In fact, the type-symmetry assumption~\eqref{eq.symmetry} can be replaced by the weaker \emph{cut-balance} condition~\cite{TsiTsi:13}, which has recently been further relaxed to a \emph{non-instantaneous} (integral) reciprocity~\cite{MartinHendrickx:2016}. Lemma~\ref{lem.converge2} can be also extended to
one-sided differential inequalities~\cite{ProCao:2017}.

Similar to the discrete-time case\footnote{As can be shown~\cite[Lemma~2.27]{RenBeardBook}, if $A(t)$ is piecewise-constant, attains values in some compact set of matrices and its consecutive switchings are separated by a positive dwell time, the model~\eqref{eq.abelson1} in fact reduces to the discrete-time model~\eqref{eq.degroot1}.}, the convergence of~\eqref{eq.abelson1} without reciprocity assumptions remains a non-trivial problem. The existing convergence criteria are mainly confined to the case where consensus of opinions is established. The most general of such criteria is a continuous-time counterpart of Lemma~\ref{lem.consensus1}, establishing exponential convergence to consensus for \emph{uniformly quasi-strong connected} (UQSC) graphs.
\begin{lem}\label{lem.consensus2}
Let $A(\cdot)$ be bounded $0\le a_{ij}(t)\le M$ and there exist $\ve,T>0$ such that the following graphs
\be\label{eq.graph-union2}
\g_{\ve,T}(t)=\g\left[\int_t^{t+T}A(s)ds\right],\quad t\ge 0
\ee
are quasi-strongly connected for any $t\ge 0$. Then the opinions in~\eqref{eq.abelson1} exponentially converge to consensus.
\end{lem}

Unfortunately, a complete proof of Lemma~\ref{lem.consensus2} is not easily available in the literature. Most of the proofs require extra assumptions, e.g. the existence of a common root node
in all the graphs $\g_{\ve,T}$~\cite{Moro:04}, piecewise-constantness of $A(t)$~\cite{LinFrancis:07,RenBeardBook,Muenz:11} or at least its continuity almost everywhere~\cite{ShiJohansson:13}. Analysis of the proofs in~\cite{LinFrancis:07,Muenz:11,ShiJohansson:13} reveals, however, the possibility to discard these additional restrictions.

The consensus criterion from Lemma~\ref{lem.consensus2} can be extended to some nonlinear consensus continuous-time algorithms~\citep{LinFrancis:07,Muenz:11,Pro13AJC,ProMatv:15,ProZhangCao:2015} and retains its validity in presence of communication delays~\cite{Muenz:11,Antonis}, whereas the validity of Lemma~\ref{lem.converge2} in presence of communication delays
seems to be a non-trivial open problem.

Whereas the uniform connectivity from Lemma~\ref{lem.consensus2} is ``almost'' necessary for consensus (being necessary for exponential consensus~\cite{LinFrancis:07} and the consensus' robustness~\cite{ShiJohansson:13}), it is only sufficient for the convergence of opinions in the time-varying Abelson model~\eqref{eq.abelson1} (for instance, we have seen~\cite{ProTempo:2017-1} that for $A(t)\equiv A$ the opinions always converge, whereas consensus requires the quasi-strong connectivity of the graph $\g[A]$).
At the same time, is can be easily shown that, similar to the discrete-time model, solutions always remain bounded since the convex hull spanned by the opinions does not expand~\cite{LinFrancis:07,TsiTsi:13,MatvPro:2013,Muenz:11}. 

%% file: 3krause.tex
\section{Opinion dynamics with bounded confidence}\label{sec.krause}

The well-known adage ``birds of a feather flock together'' prominently manifests the principle of \emph{homophily}~\citep{McPherson:2001}: similar individuals
interact more often and intensively than dissimilar people. Distancing from the members of other social groups, e.g. rejection of cultural forms they like~\cite{Mark:2003}, is an important factor of social segregation and cleavage. Humans readily assimilate opinions of like-minded individuals, accepting dissimilar opinions with discretion~\cite{Dandekar:2013}.

The idea to introduce homophily into the dynamics of opinion (``attitude'') formation has in fact been proposed by Abelson~\cite{Abelson:1964} who first realized that the time-varying model~\eqref{eq.abelson1} can reflect the effects of biased assimilation. As stated in~\cite{Abelson:1964}, the variability of
the ``contact rates'' $a_{ij}(t)$ can express that \emph{people tend to locomote into groups that share their attitudes and out of groups that do not agree with them}.
The latter phenomenon lies in the heart of many mathematical models, proposed recently and dealing with modifications of the French-DeGroot and Abelson models,
where the influence of agent $j$ on agent $i$ is the stronger, the closer are opinions of the agents~\cite{MasFlacheHelbing:2010}.
The latter principle is prominently illustrated by \emph{bounded confidence} models, attracting enormous attention of a broad research community,
from systems and control theorists to statistical physicists and data scientists.

Bounded confidence models stipulate that individuals are totally insensitive to opinions, falling outside their \emph{confidence sets}. Simple yet instructive models of this type were independently proposed by Krause~\cite{Krause} and Deffuant and Weisbuch~\cite{DeffuantWeisbuch:2000}. The Deffuant-Weisbuch model, based on the idea of gossiping, will be discussed in Section~\ref{sec.gossip}.
In this section, we are primarily concerned with the model from~\cite{Krause}, which is nowadays referred to as the \emph{Hegselmann-Krause} (HK) and has become widely known
after the publication of the work~\citep{Krause:2002}. Along with the HK model, some of its recent extensions are considered.

\subsection{The original HK model}

We start with the original model from Krause's paper~\cite{Krause}. Being an extension of the French-DeGroot model~\eqref{eq.degroot}, the HK model deals with $n$ agents, whose real opinions
$x_i\in\r$ constitute the opinion vector $x=(x_1,\ldots,x_n)\in\r^n$. We introduce the fixed \emph{range of confidence} $d>0$ and call the closed\footnote{As will be shown in the next subsection, most properties of the HK model remain valid, replacing closed confidence intervals by open ones $(x_i-d,x_i+d)$, considered e.g. in~\cite{Blondel:2009}.} set $[x_i-d,x_i+d]\subset\r$ \emph{confidence interval} of agent $i$.
Each agent $i$ ignores the opinions beyond his/her confidence interval, interacting only
with a group of \emph{trusted} individuals $I_i(x)=\{j:|x_j-x_i|\le d\}\ni\{i\}$.
Using $|I_i(x)|$ to denote their number, the $i$th agent's opinion evolves at each step as follows
\be\label{eq.hk}
x_i(k+1)=\frac{1}{|I_i(x(k))|}\sum_{j\in I_i(x(k))}x_j(k),\; i\in 1:n.
\ee
The opinion formation process~\eqref{eq.hk} is a nonlinear autonomous (time-invariant) discrete-time system
\be\label{eq.hk1}
\begin{gathered}
x(k+1)=\mathcal C(x(k))\in\r^n,\\
\mathcal C(x)=(\mathcal C_1(x),\ldots,\mathcal C_n(x))^{\top},\\
C_i(x)=\frac{1}{|I_i(x)|}\sum_{j\in I_i(x)}x_j,\quad i\in 1:n.
\end{gathered}
\ee
We refer the mapping $\mathcal C:\r^n\to\r^n$ to as the \emph{HK operator}.
On the other hand, this system can be considered as the time-varying French-DeGroot model~\eqref{eq.degroot1} with the \emph{state-dependent} matrix $W(x(k))$, where
\ben
W(x)=(w_{ij}(x)),\;\; w_{ij}(x)=\begin{cases}
1/|I_i(x)|,&j\in I_i(x)\\
0,&\text{otherwise}.
\end{cases}
\een
Introducing the corresponding influence graph
\be\label{eq.hk-graph}
\begin{gathered}
\g(x)=(V,E(x),W(x)),\\
(i,j)\in E(x)\Longleftrightarrow j\in I_i(x)\Longleftrightarrow |x_i-x_j|\le d,
\end{gathered}
\ee
one notices that the HK model stipulates the same mechanism of the opinion formation as the original
French model~\cite{French:1956}, considered in the first part~\cite{ProTempo:2017-1}.
Each agent updates its opinion to the average opinion of its neighbors
in the influence graph. The crucial difference with the French model is that this graph $\g(x)$ \emph{coevolves} with the opinions, depending on their mutual distances\footnote{The influence, or interaction graph should not be confused with the communication (information) graph, determining the agents' awareness of each other's opinions. The HK model assumes implicitly that agents are able to compute their sets $I_i(x(k))$, having thus the full information about the state vector $x(k)$. In this sense, the original HK dynamics~\eqref{eq.hk} unfolds over a social network with all-to-all communication.}. This graph may lose its connectivity, leading to disagreement of the opinions.

Dynamic networks, where the nodes and topologies have mutually dependent (coevolutionary) dynamics are actively studied by physicists~\cite{Herrera:2011}. The HK model and its modifications
constitute one important class of such networks, thoroughly studied in control theory.
In the literature, one can find many other examples of coevolutionary networks, e.g. the seminal Vicsek model of phase transitions~\cite{Vicsek}, multi-agent models of \emph{flocks}~\cite{CuckerSmale:07,Saber2006,Tanner:07} and robotic networks with range-restricted interactions~\cite{BulloBook}.

\begin{figure}[h]
\centering
    \begin{subfigure}[t]{0.48\columnwidth}
        \centering
        \includegraphics[width=\columnwidth]{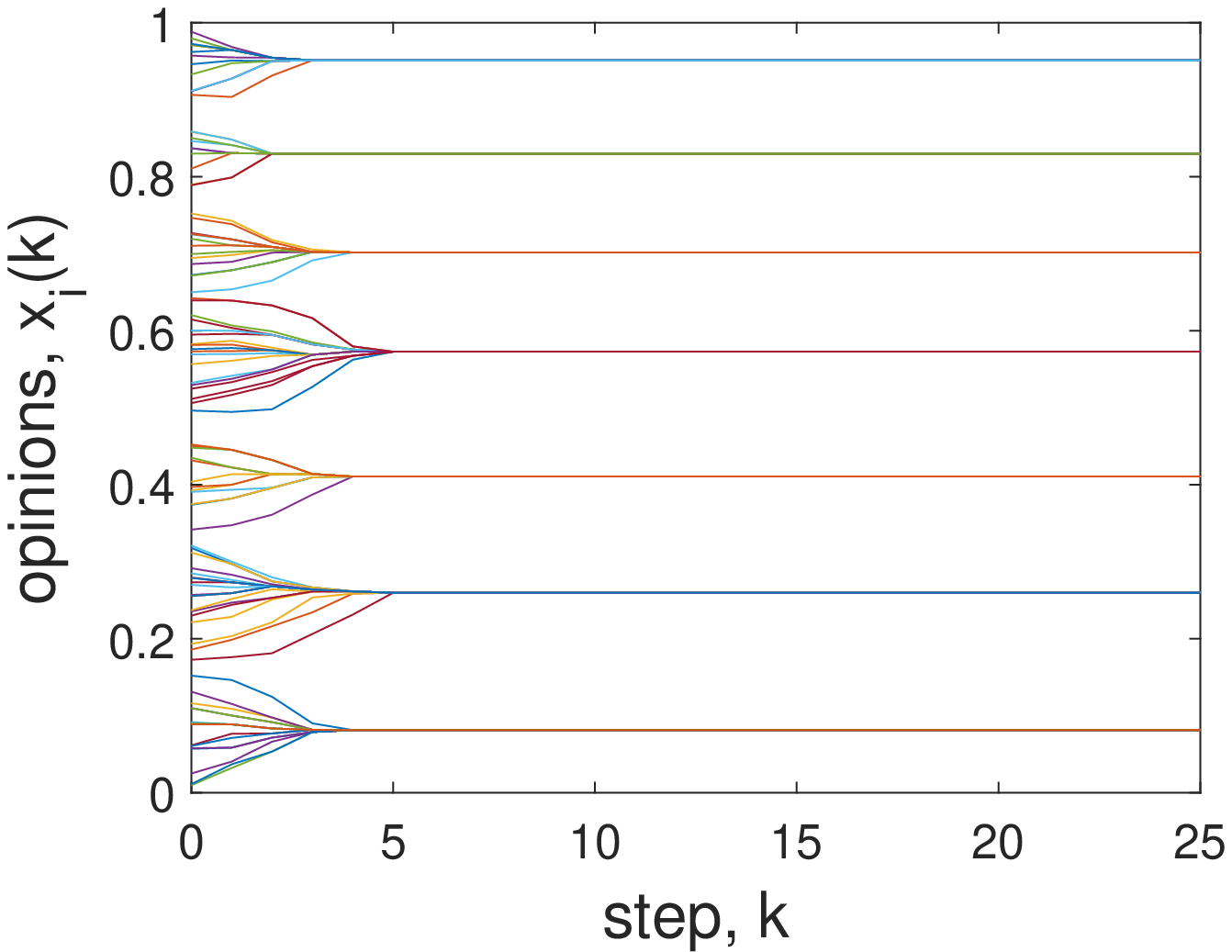}
        \caption{$d=0.05$}\label{fig.d05}
    \end{subfigure}
    \begin{subfigure}[t]{0.48\columnwidth}
        \centering
        \includegraphics[width=\columnwidth]{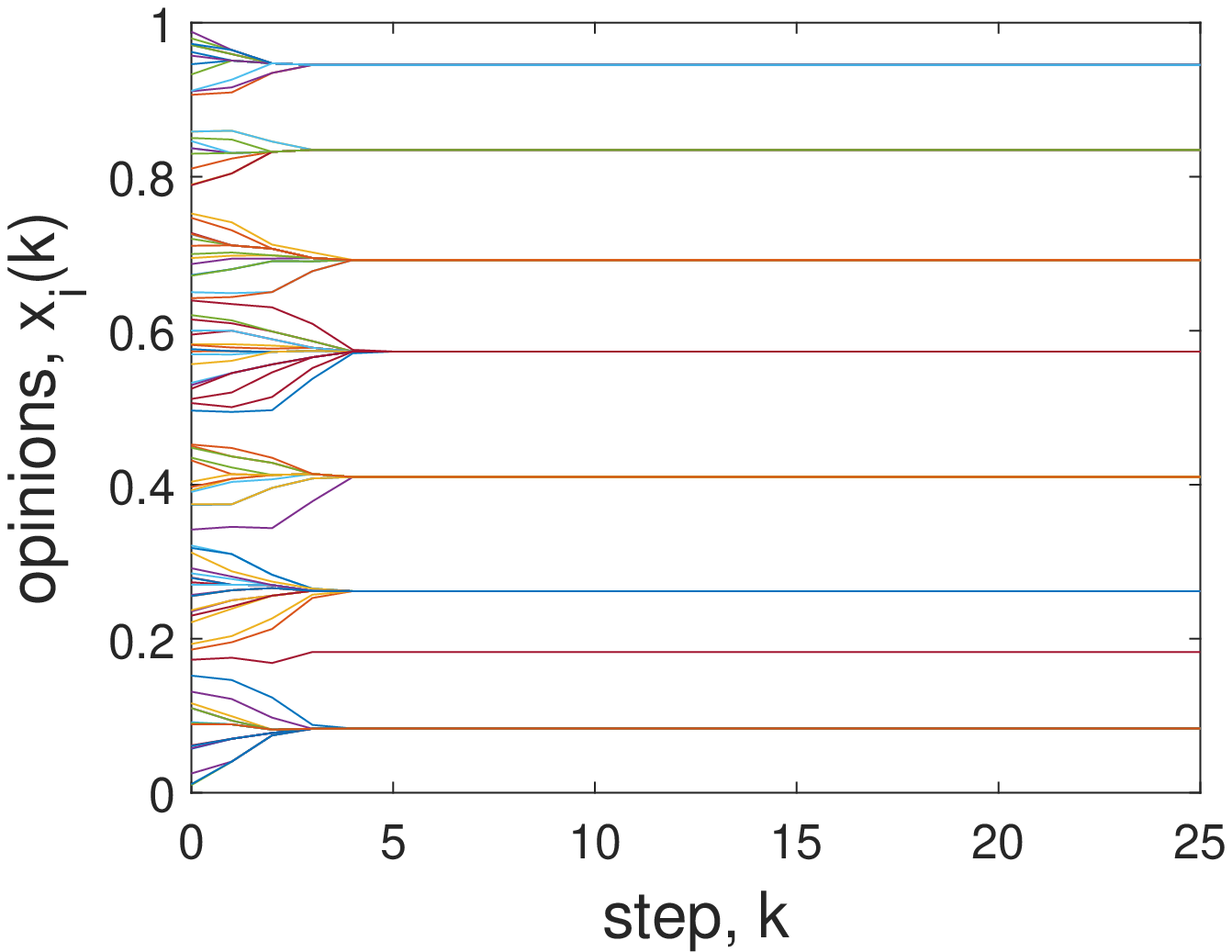}
        \caption{$d=0.06$}\label{fig.d06}
    \end{subfigure}
    \begin{subfigure}[t]{0.48\columnwidth}
        \centering
        \includegraphics[width=\columnwidth]{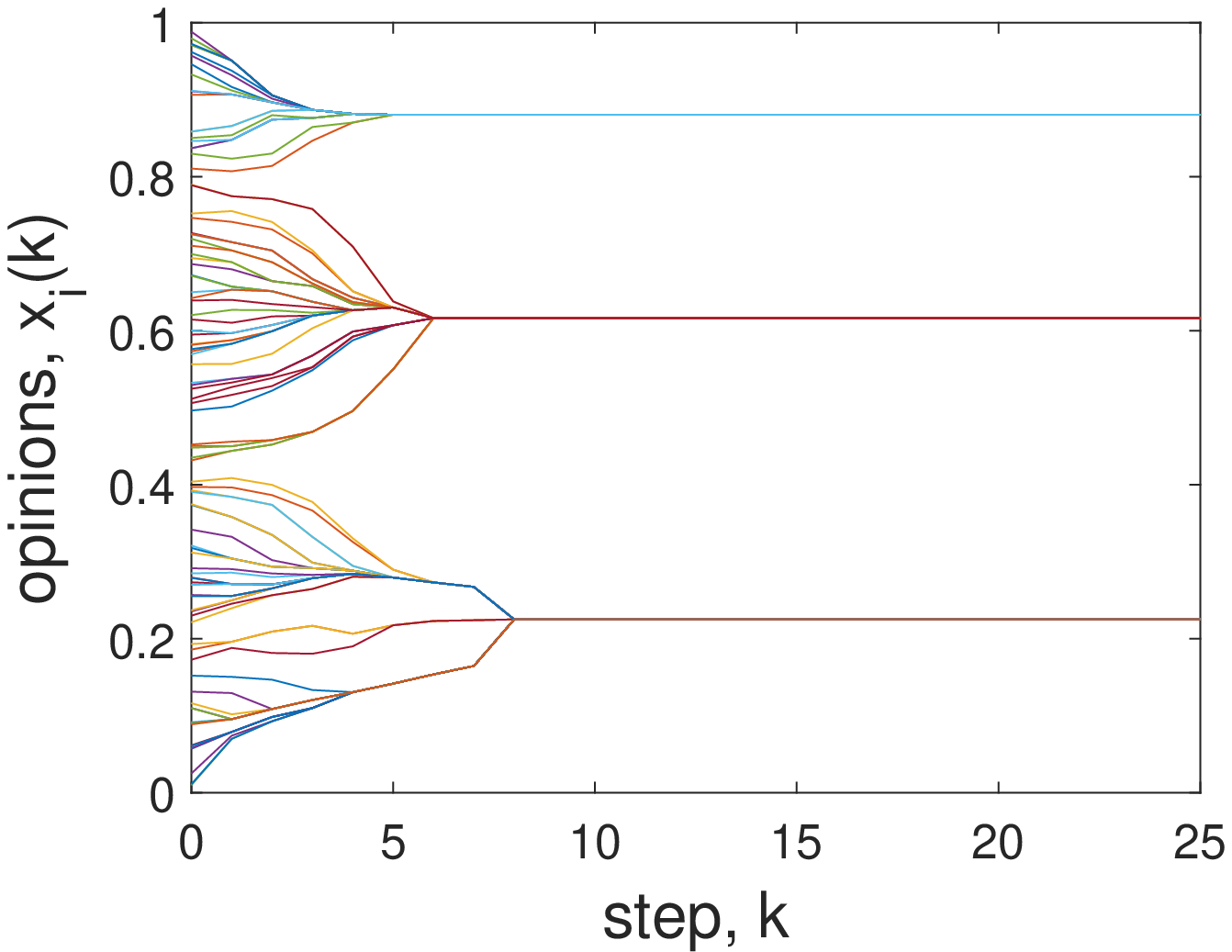}
        \caption{$d=0.11$}\label{fig.d11}
    \end{subfigure}
    \begin{subfigure}[t]{0.48\columnwidth}
        \centering
        \includegraphics[width=\columnwidth]{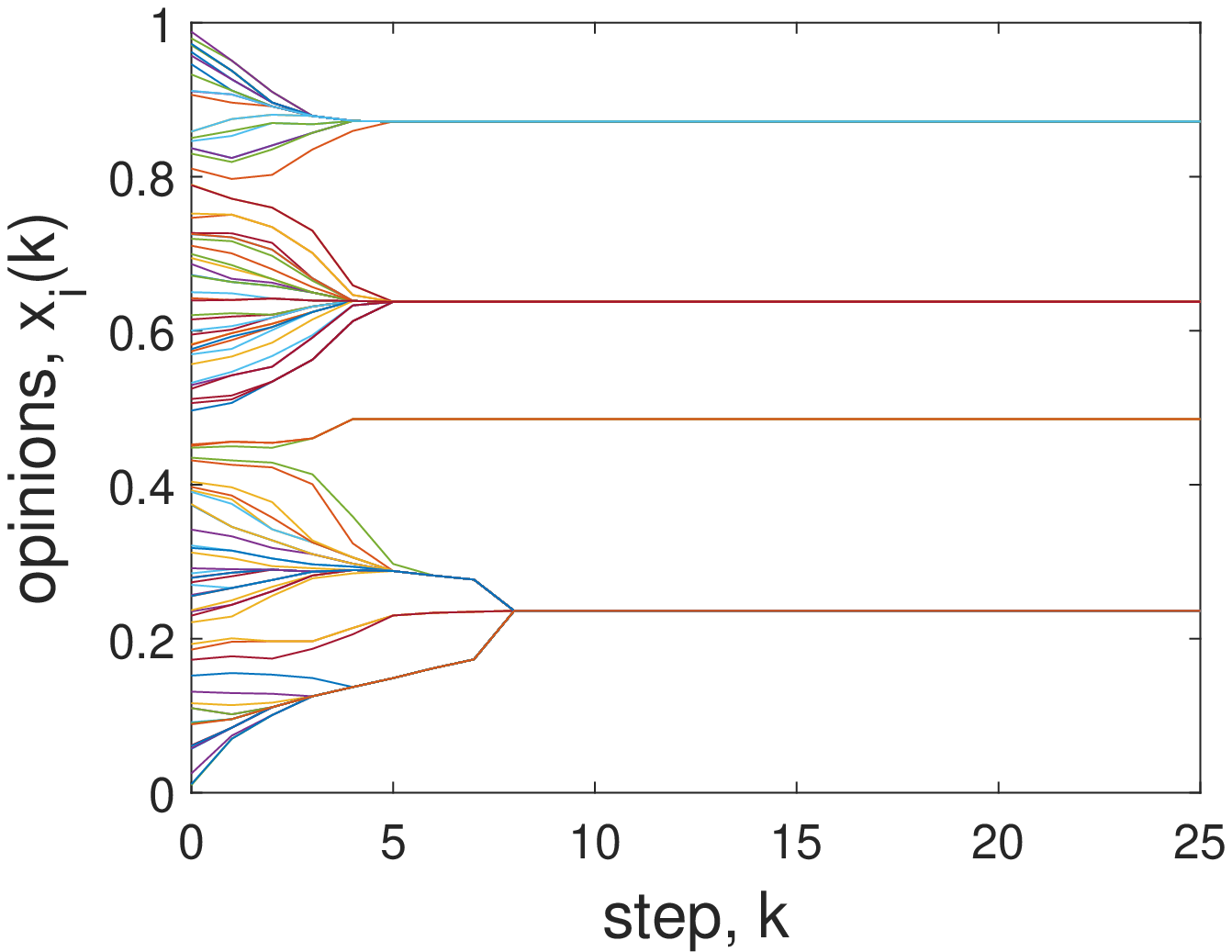}
        \caption{$d=0.12$}\label{fig.d12}
    \end{subfigure}
    \begin{subfigure}[t]{0.48\columnwidth}
        \centering
        \includegraphics[width=\columnwidth]{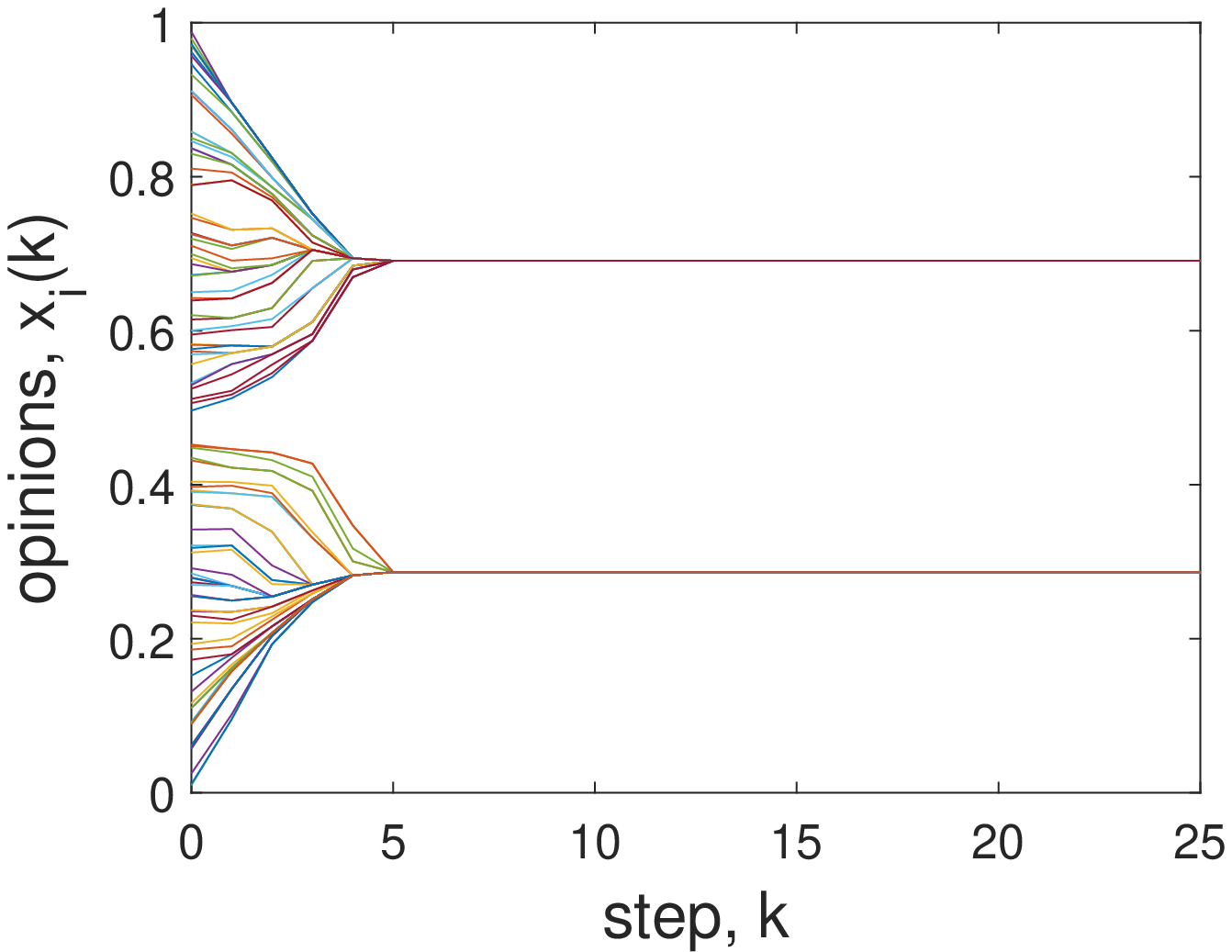}
        \caption{$d=0.2$}\label{fig.d20}
    \end{subfigure}
    \begin{subfigure}[t]{0.48\columnwidth}
        \centering
        \includegraphics[width=\columnwidth]{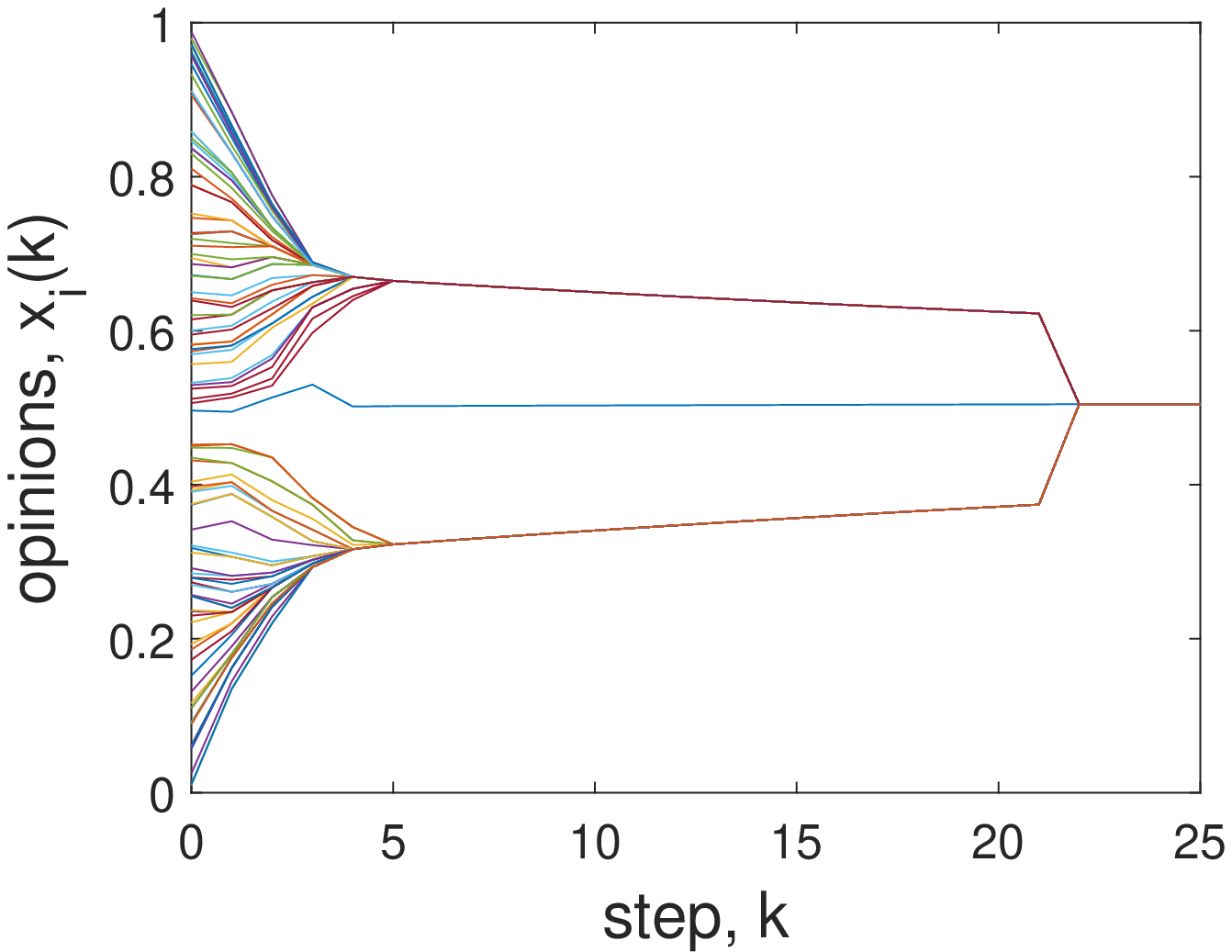}
        \caption{$d=0.25$}\label{fig.d25}
    \end{subfigure}
    \caption{The HK model with $n=100$ agents and different $d$.}\label{fig.hk}
\end{figure}

In Fig.~\ref{fig.hk}, we simulate the dynamics of $n=100$ opinions for different confidence ranges $d$.
The initial values $x_i(0)$ (same for all six experiments) are uniformly distributed on $[0,1]$.
The simulation reveals some counter-intuitive phenomena, for instance, a non-monotone dependence between $d$,
the number of clusters and the termination time. One could expect that the number of clusters is declining and the convergence time is decreasing as $d$ is growing.
In reality, an increase in $d$ can \emph{increase} the number of clusters (see the plots (a),(b) and (c),(d) in Fig.~\ref{fig.hk}).
There is no obvious dependence between $d$ and the convergence time, furthermore, for $d=0.25$ the convergence is visibly slower
than for small $d$ (Fig.~\ref{fig.hk}f) (this phenomenon of ``abnormally'' slow convergence to consensus has been reported in~\cite{Lorenz:2006}; the relevant consensus states are referred in~\cite{Lorenz:2006} as ``metastable'').
In Table~\ref{tbl.hk}, the results are compared with the prediction of a so-called \emph{$2R$-conjecture}~\cite{Blondel:2007,WangChazelle:17}, stating that for
the initial opinions, uniformly sampled from $[0,1]$ and $d=R<1/2$, the opinions converge to $\approx 1/(2R)$ clusters, separated by distances of $\approx 2R$.
\begin{table}[h]
\caption{Actual numbers of clusters vs. $2R$-conjecture~\cite{Blondel:2007}}\label{tbl.hk}
\centering 
\begin{tabular}{c | c | c} 
\hline\hline 
$d$  & Number of clusters & Rounded $1/(2d)$\\ [0.5ex] 
\hline 
0.05 & 7 & 10\\
0.06 & 8 & 8\\
0.11 & 3 & 5\\
0.12 & 4 & 4\\
0.2 & 2 & 3\\
0.25 & 1 & 2\\
\hline 
\end{tabular}
\end{table}

A natural question arises whether opinions in the HK model converge, as suggested by Fig.~\ref{fig.hk}, or can oscillate for some $x(0)$.
The following result, first proved in~\cite{Dittmer:2001}, shows that the HK model \emph{always} converges to a fixed point in a \emph{finite} number of steps.
\begin{thm}\label{thm.hk}
For any initial condition $x(0)$, the HK dynamics~\eqref{eq.hk} terminates in a finite number of steps $x(k)\equiv \bar x\,\forall k\ge k_*$, where the final opinion $\bar x$ and the termination time
depend on $x(0)$ and $d$. After the model's termination, any agents $i,j$ either are in consensus $\bar x_i=\bar x_j$ or distrust each other $|\bar x_i-\bar x_j|>d$.
\end{thm}

Note that Theorem~\ref{thm.hk} provides no information about \emph{stability} of equilibria points. In general, solutions of the HK model \emph{can} converge to unstable equilibria, however,
experiments show that for randomly chosen initial opinions such a behavior is untypical~\cite{Blondel:2009}. For criteria of (local) asymptotic stability we refer the reader to~\cite{Blondel:2009}.

There are several ways to prove Theorem~\ref{thm.hk}. The original proof~\cite{Dittmer:2001} extends the ideas from~\cite{Krause} and is based on the matrix products convergence.
Another approach is based on Lyapunov energy-like functions. The relevant methods will be discussed in the next subsections, dealing with multidimensional extensions
the HK model. The proof we outline in this subsection is based on the important \emph{order-preservation} property of the HK model~\eqref{eq.hk} and gives the best known estimate for its termination time.
\begin{lem}~\citep{Krause}\label{lem.order}
The HK operator $\mathcal C$ from~\eqref{eq.hk1} preserves the order of the elements $x_1,\ldots,x_n$, that is,
if $j_1,\ldots,j_n$ is the ordering permutation of indices
$x_{j_1}\le\ldots\le x_{j_n}$, then $\mathcal C_{j_1}(x)\le\ldots\le \mathcal C_{j_n}(x)$.
\end{lem}

Lemma~\ref{lem.order} can be proved via induction on $n$. For $n=1$, the statement is obvious. Assuming that it holds for the HK operator of dimension $n-1$, denoted $\tilde\C:\r^{n-1}\to\r^{n-1}$,
our goal is to prove it for the HK operator $\C:\r^n\to\r^n$. It suffices to consider the case where $x$ is sorted in the ascending order
$x_1\le\ldots\le x_n$, i.e. $j_i=i$. If $x_n-x_1\le d$, then $\C_1(x)=\ldots=\C_n(x)=(x_1+\ldots+x_n)/n$, and the statement is obvious.
Otherwise, let $j=\min I_n(x)<n$ and $l=\max I_1(x)>1$ and $x^{h}=(x_1,\ldots,x_{n-1})^{\top}$, $x^{t}=(x_2,\ldots,x_{n})^{\top}$ be the ``head'' and ``tail'' truncations of $x$.
For $i\le j$ one has $x_i\le x_j<x_n-d$, that is, agent $i$ is not influenced by agent $n$, and thus $\tilde\C_i(x^h)=\C_i(x)$.
For the same reason, $\tilde\C_i(x^t)=\C_i(x)$ whenever $i\ge l$. Therefore, the sequences $\{C_i(x)\}_{i=1}^j$ and $\{C_i(x)\}_{i=l}^n$ are non-decreasing.
If $l\le j$, the induction step is proved. Assuming that $j<l$, we have to show that
\be\label{eq.aux2}
C_j(x)\le C_r(x)\le C_s(x)\le C_l(x).
\ee
whenever $j<r<s<l$. Since $r,s\in I_1(x)\cap I_n(x)$,
\be\label{eq.aux2+}
\begin{gathered}
C_r(x)\overset{(*)}{=}\frac{m_r\tilde C_r(x^h)+x_n}{m_r+1}=\frac{m_r\tilde C_r(x^t)+x_1}{m_r+1},\\
C_s(x)\overset{(+)}{=}\frac{m_s\tilde C_s(x^h)+x_n}{m_s+1}\overset{(!)}{=}\frac{m_s\tilde C_s(x^t)+x_1}{m_s+1},
\end{gathered}
\ee
where $m_r=|I_r(x)|-1$ and $m_s=|I_s(x)|-1$. Recalling that $\tilde C_r(x^h)\ge \tilde C_j(x^h)=\C_j(x)$ and $x_n=\max_i x_i\ge \tilde C_j(x^h)$,
one proves the leftmost inequality in~\eqref{eq.aux2} by using the equality (*) from~\eqref{eq.aux2+}. Similarly, the equality (!) entails the rightmost inequality in~\eqref{eq.aux2}.
Using (*) and (+), the mid inequality in~\eqref{eq.aux2} shapes into
\[
m_r\tilde C_r(x^h)\le m_s \tilde C_s(x^h)+(m_r-m_s)x_n.
\]
To prove the latter inequality, note that $\tilde C_s(x^h)\ge \tilde C_r(x^h)$ by assumption, $x_n=\max_i x_i\ge \tilde C_r(x^h)$ and $I_s(x)\subseteq I_r(x)$, entailing that $m_r\ge m_s$.
This finishes the proof of~\eqref{eq.aux2} and of the induction step.$\blacksquare$

\begin{figure}[h]
\centering
\includegraphics[width=0.75\columnwidth]{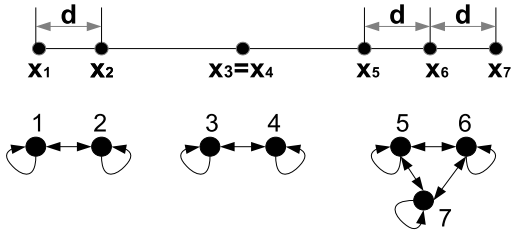}
\caption{Opinions of $n=7$ agents and the graph $\g(x)$}\label{fig.hk-graph}
\end{figure}

From now on until the end of this subsection, the agents' opinions are numbered in the ascending order
\be\label{eq.ascending}
x_1\le\ldots\le x_n.
\ee
If the initial vector of opinions $x(0)$ is sorted as in~\eqref{eq.ascending}, this order of opinions is preserved at any iteration due to Lemma~\ref{lem.order}.
We say that the opinions $(x_i,\ldots,x_{m})$ constitute a \emph{$d$-chain}~\cite{Dittmer:2001}
if the distances between consecutive opinions $x_{i+1}-x_j,\ldots,x_{m}-x_{m-1}$ are $\le d$, that is, the graph $\g(x)$ from~\eqref{eq.hk-graph} contains a chain of arcs
$i\leftrightarrow i+1\leftrightarrow\ldots\leftrightarrow i+m$. Obviously, the vector of opinions $x$ consists of several \emph{maximal} $d$-chains (that are not contained
by any longer $d$-chain), which correspond to the \emph{connected components} of the graph $\g(x)$. This is illustrated in Fig.~\ref{fig.hk-graph}, where the opinions split into three maximal $d$-chains $(x_1,x_2)$, $(x_3,x_4)$ and $(x_5,x_6,x_7)$, standing for the three connected components of the graph $\g(x)$.

It can be easily shown that two different maximal $d$-chains can never merge, and the corresponding sets of agents
do not influence each other at any step.
\begin{lem}\label{lem.d-sep}
Suppose that the initial opinions $x_i(0)$ are sorted in the ascending order~\eqref{eq.ascending} and $x_{i+1}(0)-x_i(0)>d$. Then $x_{i+1}(k)$ is non-decreasing and $x_i(k)$ is non-increasing in $k$, and thus
$x_{i+1}(k)-x_i(k)>d$. In other words, two maximal $d$-chains cannot merge.
\end{lem}

Proof. We will show via induction on $k$ that
\be\label{eq.aux3}
x_{i}(k+1)\le x_i(k)\le x_{i+1}(k)-d\le x_{i+1}(k+1)-d.
\ee
We prove the induction base $k=0$. Agent $i$ can interact at step $k=0$ only with some of agents $1,\ldots,i-1$, and thus $x_i(1)\le x_i(0)=\max_{j\le i}x_j(0)$. Similarly, agent $i+1$ can interact only with agents $i+2,\ldots,n$, so that $x_{i+1}(1)\ge x_{i+1}(0)$. This proves~\eqref{eq.aux3} for $k=0$. The step from $k$ to $k+1$ is proved in the same way.$\blacksquare$
\begin{cor}\label{cor.d-sep1}
If two agents $i$ and $j$ belong to different connected components of $\g(x(k_0))$, there are no walks connecting them in any of the graphs $\g(x(k))$, $k\ge k_0$.
As $k$ grows, the strong components of $\g(x(k))$ can split into smaller components but cannot merge.
\end{cor}

Lemma~\ref{lem.d-sep} can also be reformulated as follows: in each maximal $d$-chain $x_j(k)\le\ldots\le x_{m}(k)$
the leftmost opinion $x_j(k)$ is non-decreasing $x_j(k+1)\ge x_j(k)$, whereas the rightmost opinion is non-increasing
$x_{m}(k+1)\le x_m(k)$. In particular, the \emph{diameter} of the $d$-chain $x_m(k)-x_j(k)$ is non-increasing.

If the diameter of a maximal $d$-chain is not greater than $d$, at the next step this chain collapses into a group of identical opinions $x_i=x_{i+1}=\ldots=x_{i+m}$,
which we henceforth refer to as a $x_i=x_{i+1}=\ldots=x_{i+m}$ (an example of such a cluster is the pair of opinions $x_3,x_4$ in Fig.~\ref{fig.hk-graph}).
This happens e.g. with the maximal $d$-chain with only two opinions $x_i(k)<x_{i+1}(k)$. Maximal $d$-chains containing $3$ or $4$ opinions in fact also collapse into consensus clusters after, respectively, $2$ and $5$ steps~\cite{Krause}, and hence the HK model with $n<5$ agents always converges to consensus.
This statement does not hold for $n>5$: maximal $d$-chains with $5$ and more opinions can split into shorter $d$-chains,
which in turn can further split or converge to different consensus clusters. For this reason, the HK model with $n\ge 5$ agents may fail to reach consensus even when $\g(x(0))$ is connected.

A more accurate analysis of $d$-chains reveals the following important property~\cite{BhatChazelle:2013}.
\begin{lem}\label{lem.bhat}
During each two consecutive steps $k$ and $k+1$, any maximal chain in the vector $x(k)$
collapses into a singleton, splits into several maximal $d$-chains or reduces in diameter by at least $d/(n^2)$.
\end{lem}

There can be at most $n-1$ times $k$ at which one of the chains collapses, and at most $n-1$ splitting times.
Obviously, the sum of the diameters of all $d$-chains is not greater than $(n-1)d$, so the diameter can be decreased no more than $(n-1)n^2$ times. 
Hence, the HK dynamics terminates in no more than $k_*\le 2((n-1)n^2+2(n-1))=2n^3-2(n-1)^2$ steps.
\begin{cor}\cite{BhatChazelle:2013}\label{cor.n3}
The HK model with $n$ agents terminates in no more that $O(n^3)$ steps.
\end{cor}

The polynomial convergence time has been first conjectured in~\cite{Chazelle:11}, where the HK model has been proved to terminate in $n^{O(n)}$ steps.
An alternative proof of Corollary~\ref{cor.n3}, based on Lyapunov analysis, has been given in~\cite{MohajerTouri:2013} (with the upper bound of termination time $3n^3+n$). 
More conservative polynomial estimates for the termination time have been obtained in~\cite{EtesamiBasar:2015,BulloBook,TouriNedic:2011}; the approaches developed there
are also applicable to more general \emph{multidimensional} HK models, considered in the next subsection. Notice that Lemma~\ref{lem.bhat} gives only an \emph{upper} bound of the convergence time. To the best of the authors' knowledge, the tightness of this estimate remains a non-trivial open problem. For some initial conditions
the HK model terminates in $\Omega(n^2)$ steps~\cite{WedinHegarty:2015-1}.

\subsection{The multidimensional HK model}

A natural extension of the HK model deals with \emph{multidimensional} opinions $x_i(k)\in\r^m$~\cite{Nedic:2012}. Choosing some norm $\|\cdot\|$ on $\r^m$, the confidence interval for agent $i$ is  replaced by the \emph{ball} $\{\xi\in\r^m:\|\xi-x_i\|\le d\}$ and hence the set of trusted individuals is defined as
\be\label{eq.trust-multidim}
I_i(x)=\{j:\|x_j-x_i\|\le d\},
\ee
where $x\in\r^{nm}$ denotes the column vector, obtained by stacking $x_1,\ldots,x_n\in\r^m$ on top of each other.
Here $\|\cdot\|$ can be an arbitrary norm on $\r^m$, however most of the existing works~\cite{Nedic:2012,BhatChazelle:2013,Etesami:2013,EtesamiBasar:2015,Martinsson:16} deal with the Euclidean norm $\|\xi\|=\sqrt{\xi^{\top}\xi}$.

Considering the scalar elements of the multidimensional opinions as individual's positions on different issues, the definition of trust sets~\eqref{eq.trust-multidim} imposes
an implicit \emph{dependence} between these issues. In particular, two individuals $i,j$ that strongly disagree on the $s$th issue (e.g. $|x_{i,s}-x_{j,s}|\ge d$)
ignore each other's positions on all remaining issues since $i\not\in I_j(x),j\not\in I_i(x)$.

\begin{figure}[h]
\centering
\begin{subfigure}[t]{0.49\columnwidth}
        \centering
        \includegraphics[width=\columnwidth]{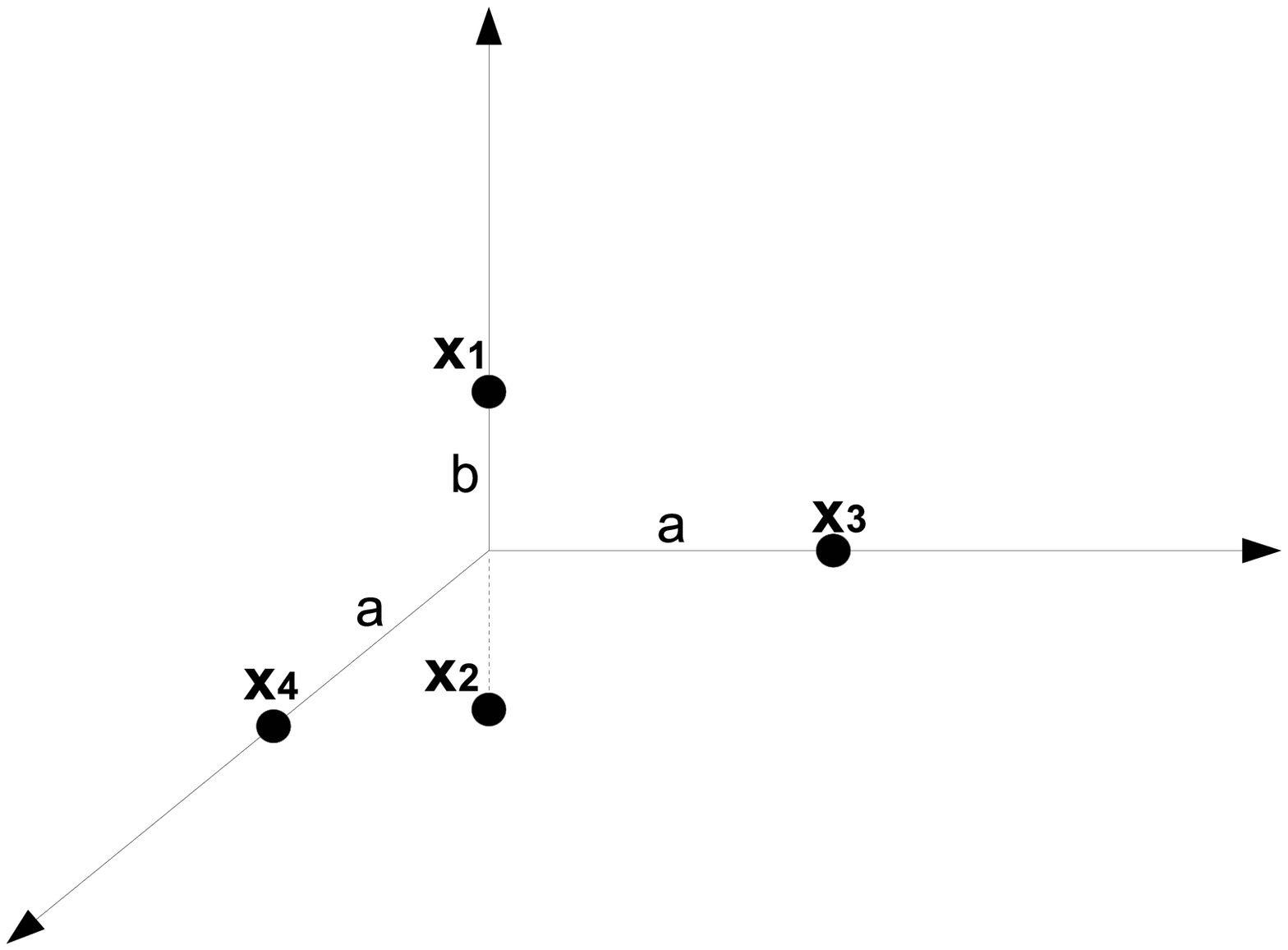}
        \caption{Opinions $x_i(0)$}
    \end{subfigure}\hfill
    \begin{subfigure}[t]{0.49\columnwidth}
        \centering
        \includegraphics[width=\columnwidth]{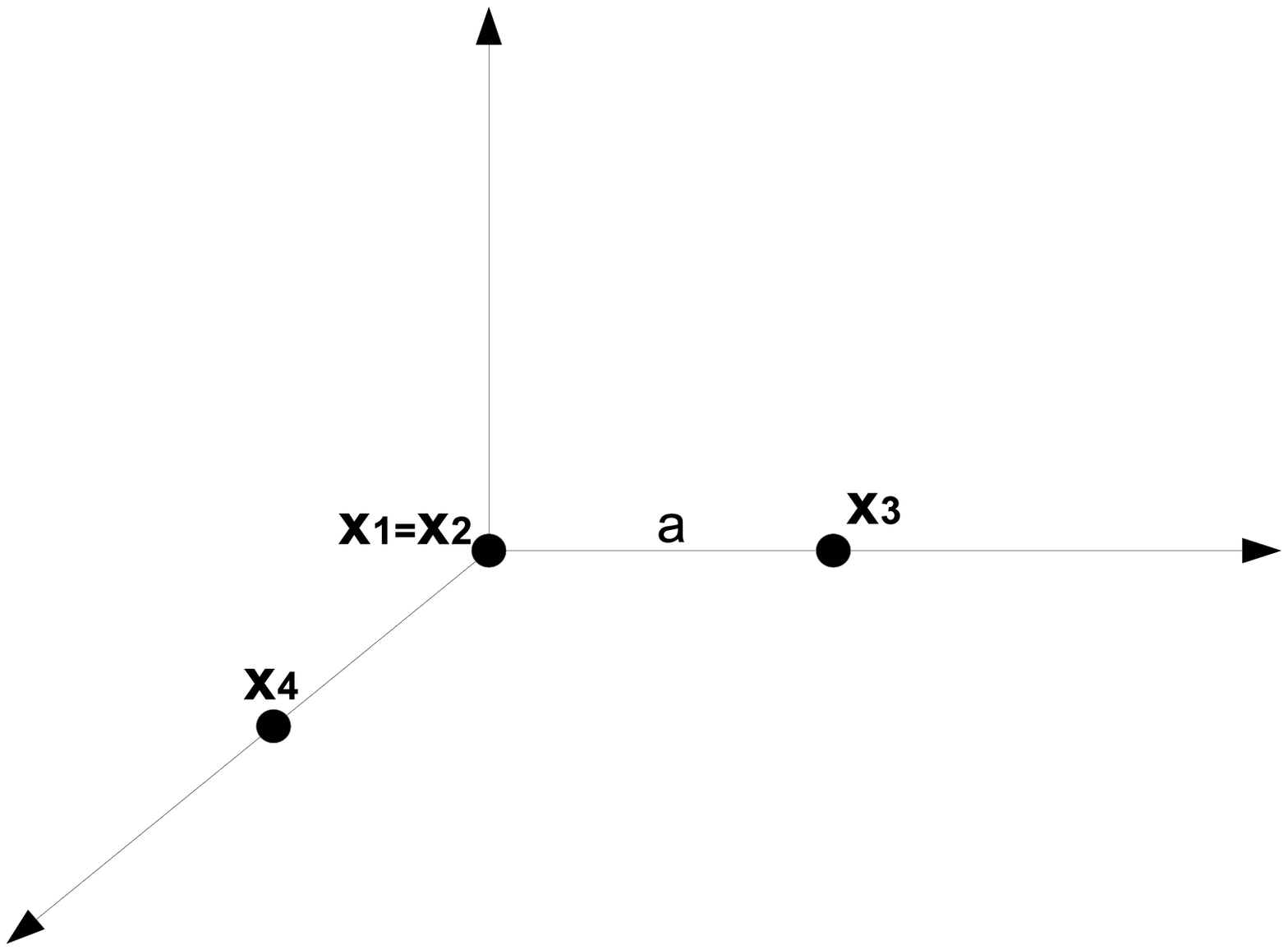}
        \caption{Opinions $x_i(1)$}
    \end{subfigure}
\begin{subfigure}[t]{0.49\columnwidth}
        \centering
        \includegraphics[width=0.3\columnwidth]{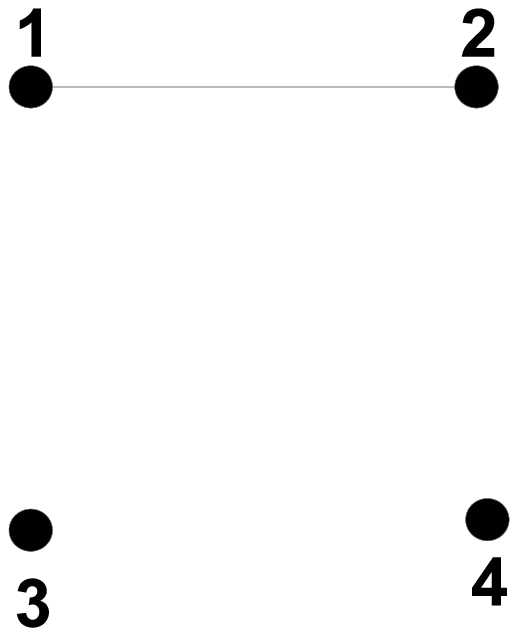}
        \caption{The graph $\g(x(0))$}
    \end{subfigure}\hfill
    \begin{subfigure}[t]{0.49\columnwidth}
        \centering
        \includegraphics[width=0.3\columnwidth]{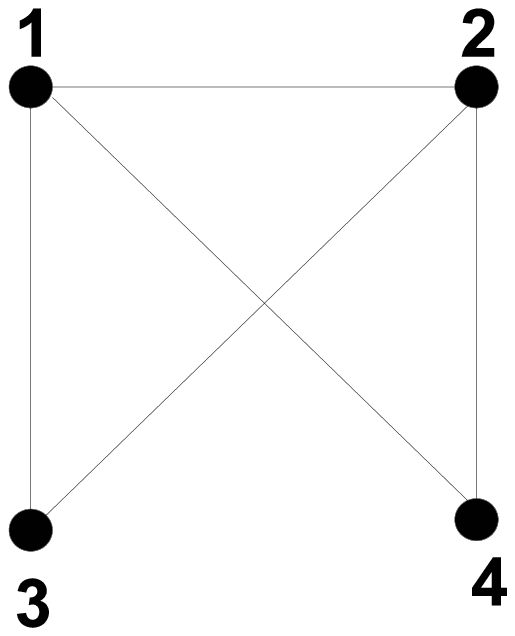}
        \caption{The graph $\g(x(1))$}
    \end{subfigure}
\caption{Two connected components of $\g(x)$ merge}\label{fig.hk-mult1}
\end{figure}

Unlike the scalar case, for $m>1$ the connected components of the graph $\g(x)$ can not only split, but also merge as shown in Fig.~\ref{fig.hk-mult1}. Consider $n=4$ opinion vectors $x_i(0)\in\r^3$, being the vertices of a tetrahedron $x_1(0)=(0,0,b)$, $x_2(0)=(0,0,-b)$, $x_3(0)=(a,0,0)$, $x_4(0)=(0,a,0)$, where $0<b<d/2$ and $\sqrt{d^2-b^2}<a\le d$.
It can be easily shown that the graph $\g(x(0))$ has three connected components (Fig.~\ref{fig.hk-mult1}c) since $I_1(x)=I_2(x)=\{1,2\}$ and $I_i(x)=\{i\}$ for $i=3,4$.
At the next step (Fig.~\ref{fig.hk-mult1}b) one has
\[
\begin{gathered}
x_1(1)=x_2(1)=(0,0,0),\; x_3(1)=x_3(0),x_4(1)=x_4(0),
\end{gathered}
\]
and thus the graph $\g(x(1))$ is \emph{connected} (Fig.~\ref{fig.hk-mult1}d) (in fact, agents reach consensus in $k=3$ steps).

A natural question thus arises whether result of Theorem~\ref{thm.hk} holds for the multidimensional HK model~\eqref{eq.hk},~\eqref{eq.trust-multidim}, that is,
the dynamics terminate in finite time. An affirmative answer is giving by the following theorem.
\begin{thm}\label{thm.hk-mult}
For any choice of the norm $\|\cdot\|$, the model~\eqref{eq.hk},\eqref{eq.trust-multidim} terminates in finite number of steps.
\end{thm}

The simplest way to prove Theorem~\ref{thm.hk-mult} is to apply Lemma~\ref{lem.converge1}. It can be easily shown that the HK dynamics~\eqref{eq.hk},~\eqref{eq.trust-multidim} can be written as the time-varying French-DeGroot model~\eqref{eq.degroot1}, where the stochastic matrix $W(k)=W(x(k))$ is \emph{state-dependent} and satisfies all assumptions of Lemma~\ref{lem.converge1}. Hence the opinions converge, and it remains to prove \emph{finite-time} convergence. Considering the graph of persistent interactions $\g_{\infty}$ corresponding to some specific solution.
As we have noticed, condition (a) in Lemma~\ref{lem.converge1} implies that if nodes $i$ and $j$ are not connected in $\g_{\infty}$,
they are not connected in $\g(x(k))$ for large $k$. Hence for large $k$ the nodes from different components of $\g_{\infty}$ do not influence each other. Consider now connected components $\g_{\infty}^s=(V^s,E_{\infty}^s)$ of the graph $\g_{\infty}$. Thanks to Lemma~\ref{lem.converge1}, in each component consensus is established
\[
x_i(k)\xrightarrow[k\to\infty]{}\bar x^s\quad\forall i\in V^s.
\]
Therefore, for large $k$ one has $|x_i(k)-x_j(k)|<d\,\forall i,j\in V^s$, and hence $I_i(x)=V^s\,\forall i\in V^s$.
This means that at some step $k=k_0$ the opinions in each connected component of $\g(x(k))$ become equal
\[
x_i(k_0+1)=\frac{1}{|V^s|}\sum_{j\in V^s}x_j(k_0)\quad\forall i\in V^s,
\]
that is, the HK model terminates in finite time. $\square$

A natural question arises whether the termination time depends on the number of agents $n$ polynomially. The positive answer has been given in~\cite{Etesami:2013,EtesamiBasar:2015,BhatChazelle:2013,TouriPhD}. In the Euclidean norm, the best existing estimate for the dimensions $m\ge 2$ seems to be $2n^4$ steps~\cite{Martinsson:16}.
This and other existing estimates for the convergence time of the multidimensional HK model are based on special Lyapunov functions that will be discussed in the next subsection.
It is remarkable that the best known \emph{lower} bound for the dimensions $m\ge 2$ is $\Omega(n^2)$ (same as in one-dimensional case); as demonstrated in~\cite{BhatChazelle:2013}, the agents placed in the vertices of regular (planar) $n$-gon with the side $d$ reach consensus in no less than $n^2/28$ steps.

It can be noticed that Theorem~\ref{thm.hk-mult} remains valid for more general distance-based models of opinion formation,
examined in~\cite{JabinMotsch:2014,MotschTadmor:2013}. Given a function $\vp:[0,\infty)\to [0,\infty)$, the multidimensional opinion vectors $x_i(k)\in\r^m$ evolve as follows
\be\label{eq.jabin}
\begin{gathered}
x_i(k+1)=\frac{\sum_{j=1}^n\vp_{ij}(k)x_j(k)}{\sum_{j=1}^N\vp_{ij}(k)},\\ \vp_{ij}(k):=\vp(|x_j(k)-x_i(k)|^2).
\end{gathered}
\ee
Obviously, the multidimensional HK model with Euclidean norm is a special case of the model~\eqref{eq.jabin}, where $\vp$ stands for the indicator function of the interval $[0,d^2]$. In~\cite{MotschTadmor:2013}, another function was considered
\[
\vp(\sigma)=
\begin{cases}
a,\quad \sigma\le d_1^2\\
b, \quad \sigma\in (d_1^2,d_2^2),\\
0, \quad\sigma>d_2^2,
\end{cases}
\]
where $0<a<b$ and $0<d_1<d_2$. This function $\vp(\sigma)$ represents the phenomenon of \emph{heterophily}: moderately
distant opinions are attracted more intensively than similar ones. A counter-intuitive phenomenon, reported
in~\cite{MotschTadmor:2013}, is the facilitation of consensus by these ``heterophilous'' interactions.

The result of Theorem~\ref{thm.hk-mult} (except for the finite-time convergence) can be easily extended to systems~\eqref{eq.jabin} with special functions $\vp(\cdot)$.
\begin{lem}\label{thm.hk-mult1}
Suppose that $\vp(\sigma)\in\{0\}\cup [a,b]\,\forall\sigma\ge 0$ for some constants $0<a<b$ and, furthermore, $\vp(0)>0$.
Then the opinions, obeying~\eqref{eq.jabin}, converge (yet do not stabilize after finite number of steps).
\end{lem}

Lemma~\ref{thm.hk-mult1} easily follows from Lemma~\ref{lem.converge1}. As will be discussed in the next subsection, in fact the system~\eqref{eq.jabin} converges for many other functions $\vp(\cdot)$, including \emph{continuous} ones; the relevant convergence results are based on special Lyapunov functions. As a further extension of~\eqref{eq.jabin}, one may consider a model where each pair of agents $(i,j)$ is endowed with its own distance-measuring function $\vp^{ij}(\sigma)$; without loss of generality we assume that
$\vp^{ii}(\sigma)\equiv \vp^{ii}>0$.
\be\label{eq.jabin1}
\begin{gathered}
x_i(k+1)=\frac{\sum_{j=1}^n\vp_{ij}(k)x_j(k)}{\sum_{j=1}^N\vp_{ij}(k)},\\ \vp_{ij}(k):=\vp^{ij}(|x_j(k)-x_i(k)|^2).
\end{gathered}
\ee
Obviously, if $\vp^{ij}(\sigma)\in\{0\}\cup [a,b]$ for some $0<a<b$ and $\vp^{ij}(\sigma)>0\Leftrightarrow \vp^{ji}(\sigma)>0$, the result of Lemma~\ref{thm.hk-mult1} remains valid for the heterogeneous model~\eqref{eq.jabin1}. A special case of model~\eqref{eq.jabin1} has been proposed in~\cite{Blondel:2009}, choosing the mappings $\vp^{ij}$ as follows
\[
\vp^{ij}(\sigma)=\begin{cases}
w_j,\quad &\sigma<d^2\\
0,\quad&\sigma\ge d^2.
\end{cases}
\]
Here $w_1,\ldots,w_n>0$ are positive ``weights'' or ``reputations''~\cite{ChenGlass:16} of the agents. Similarly to the usual HK model, in the scalar case ($m=1$) such a model provides the order preservation of the opinions~\cite{Blondel:2009}.

\subsection{Lyapunov methods for the HK model}

Henceforth the ``term HK model'' stands for the multidimensional model~\eqref{eq.hk},\eqref{eq.trust-multidim} with the \emph{Euclidean} norm $\|x\|=|x|$.

The standard Lyapunov function used to study consensus algorithms~\eqref{eq.degroot1},\eqref{eq.abelson1} is the diameter of the convex hull, spanned by the agents' opinions~\cite{Moro:05,LinFrancis:07}. This Lyapunov function, however, appears to be most useful in the case where consensus of opinions is established (and thus their convex hull collapses into a singleton), whereas the opinions in the HK model, in general, split into several clusters. The special structure of the HK model~\eqref{eq.hk}, however, implies the existence of another piecewise-smooth Lyapunov function
\be\label{eq.energy}
\mathcal E(x)\dfb\sum_{i,j=1}^n\min(|x_i-x_j|^2,d^2).
\ee
This ``energy'' function is a special case of more general Lyapunov functions, proposed in the seminal paper~\cite{Roozbehani:2008} to examine some types of nonlinear consensus algorithms (a similar Lyapunov function has been also used to study ``continuum'' bounded confidence opinion dynamics in~\cite{Blondel:2007}).
The function ~\eqref{eq.energy} proves to be non-increasing along the system's trajectories; moreover, it strictly decreases until the opinion evolution terminates,
as implied by the following.
\begin{prop}\cite{Roozbehani:2008,BhatChazelle:2013}\label{prop.energy}
For any solution of the HK model and $k=0,1,\ldots$, the inequality holds
\be\label{eq.energy-decrease}
\mathcal E(x(k))-\mathcal E(x(k+1))\ge 4\sum_{i=1}^n|x_i(k+1)-x_i(k)|^2.
\ee
\end{prop}

The inequality~\eqref{eq.energy-decrease} implies the following bound for kinetic $2$-energy (defined in~\eqref{eq.s-energy}) of the HK model
\be\label{eq.k2}
K(2)\le \mathcal E(x(0))\le d^2n(n-1).
\ee
Using elegant techniques from algebraic graph theory, it has been shown in~\cite{Martinsson:16} that during each step of the opinion formation either two opinions merge or the energy $\mathcal E(x(k))$ is decreased by at least $d/(2n^2)$, and hence the HK dynamics terminates in $O(n^4)$ steps independent of the opinions' dimension $m$.

Proposition~\eqref{prop.energy} can be extended to a more general model~\eqref{eq.jabin} with a non-increasing function $\vp(\cdot)$.
\begin{lem}\cite{JabinMotsch:2014}\label{lem.energy}
Suppose that $\vp(\cdot)$ in~\eqref{eq.jabin} is non-increasing. Then the ``energy'' function
\be\label{eq.energy+}
\mathcal E_{\vp}(x)=\sum_{i,j=1}^n\Phi(|x_i-x_j|^2),\; \Phi(r)=\int_0^r\vp(\sigma)d\sigma,
\ee
is non-increasing and satisfies the inequality
\be\label{eq.energy-decrease+}
\begin{aligned}
\mathcal E_{\vp}&(x(k))-\mathcal E_{\vp}(x(k+1))\geq \\
&\geq\sum_{i,j=1}^n\vp_{ij}(k)|\Delta x_i(k)+\Delta x_j(k)|^2\\
&\geq 4\vp(0)\sum_{i=1}^n|\Delta x_i(k)|^2,
\end{aligned}
\ee
where $\Delta x_i(k)=x_i(k+1)-x_i(k)$.
\end{lem}
Obviously, Lemma~\ref{lem.energy} implies Proposition~\ref{prop.energy} since in the case of the usual HK model one has $\Phi(r)=\min(r,d^2)$ and $\vp(0)=1$. We give the sketch of the proof of Lemma~\ref{lem.energy}, presented\footnote{In fact, Proposition~4.1 in~\cite{JabinMotsch:2014} reports a stronger inequality $\mathcal E(x(k))-\mathcal E(x(k+1))\geq 4\sum_{i,j}\vp_{ij}(k)|\Delta x_i(k)|^2$ whose proof seems to be elusive: the latter inequality relies on~\eqref{eq.energy-decrease+} and the incorrect inequality
$-\sum_{i,j}\vp_{ij}(k)|\Delta x_i(k)+\Delta x_j(k)|^2\leq -4\sum_{i,j}\vp_{ij}(k)|\Delta x_i(k)|^2$ (which holds with $\geq$ instead of $\leq$).} in~\cite{JabinMotsch:2014} (Proposition 4.1).
The proof is based on the following three relations. First, the function $\Phi(r)$ is \emph{concave} ($\Phi'(r)=\vp(r)$ is non-increasing) and hence
\be\label{eq.aux-concave}
\Phi(a)-\Phi(b)\ge \vp(a)(a-b)\quad\forall a,b\ge 0.
\ee
Second, $\vp_{ij}(k)=\vp_{ji}(k)$, and therefore
\be\label{eq.aux-symm}
-2\sum_{i,j}\vp_{ij}(k)\xi_i^{\top}(\eta_j-\eta_i)=\sum_{i,j}\vp_{ij}(k)(\xi_j-\xi_i)^{\top}(\eta_j-\eta_i)
\ee
for any set of vectors $\xi_i,\eta_i\in\r^m$ (here $i=1,\ldots,n$). Finally, for each $i$~\eqref{eq.jabin} implies the following
\be\label{eq.aux-delta}
\sum_{j=1}^n\vp_{ij}(k)\Delta x_i(k)=\sum_{j=1}^n\vp_{ij}(k)(x_j(k)-x_i(k)).
\ee
Denoting $z_{ij}(k)=x_j(k)-x_i(k)$, one obtains
\be\label{eq.aux-aux}
\begin{aligned}
&\sum_{i,j}\vp_{ij}(k)(\Delta x_j(k)-\Delta x_i(k))^{\top}z_{ij}(k)\overset{\eqref{eq.aux-symm}}{=}\\
&=-2\sum_{i,j}\vp_{ij}(k)\Delta x_i(k)^{\top}z_{ij}(k)\overset{\eqref{eq.aux-delta}}{=}\\
&=-2\sum_{i,j}\vp_{ij}(k)|\Delta x_i(k)|^2=\\&=-\sum_{i,j}\vp_{ij}(k)\left(|\Delta x_j(k)|^2+|\Delta x_i(k)|^2\right).
\end{aligned}
\ee
By noticing that $z_{ij}(k+1)=z_{ij}(k)+\Delta x_j(k)-\Delta x_i(k)$, the latter equality entails that
\[
\begin{split}
\mathcal E_{\vp}&(x(k))-\mathcal E_{\vp}(x(k+1))\overset{\eqref{eq.energy+},\eqref{eq.aux-concave}}{\geq} \\
&\geq\sum_{i,j}\vp_{ij}(k)(|z_{ij}(k)|^2-|z_{ij}(k+1)|^2)=\\
&=-\sum_{i,j}\vp_{ij}(k)|\Delta x_j(k)-\Delta x_i(k)|^2-\\
&-2\sum_{i,j}\vp_{ij}(k)(\Delta x_j(k)-\Delta x_i(k))^{\top}z_{ij}(k)\overset{\eqref{eq.aux-aux}}{=}\\
&=\sum_{i,j}\vp_{ij}(k)|\Delta x_j(k)+\Delta x_i(k)|^2\geq\\
&\geq 4\sum_{i}\vp_{ii}(k)|\Delta x_i(k)|^2=4\vp(0)\sum_{i}|\Delta x_i(k)|^2.\square
\end{split}
\]

Using the inequality~\eqref{eq.energy-decrease+}, it is possible to establish convergence of the model~\eqref{eq.jabin} when the function $\vp(\cdot)$ does not satisfy the conditions of Lemma~\ref{thm.hk-mult1}~\cite{Roozbehani:2008,MotschTadmor:2013,JabinMotsch:2014}. The result of~\cite[Theorem~3]{JabinMotsch:2014} establishes convergence of the model~\eqref{eq.jabin} for any non-increasing and \emph{concave} function $\vp(\cdot)$ with a compact support, on which the inequality $|\vp'(r)|^2\le C\vp(r)$ should hold for some constant $C>0$. The
method developed in~\cite{Roozbehani:2008} allows to discard the concavity assumption~\cite[Corollary~1]{Roozbehani:2008} in the scalar case.

It should be noticed that~\eqref{eq.energy} is not the only Lyapunov function that can be used to examine the HK model. Alternative Lyapunov functions have been employed  in ~\cite{MohajerTouri:2013,CoulsonTouri:2015} and~\cite{TouriNedic:2011,Nedic:2012} (the latter works deal with a special Lyapunov \emph{functional}, based on the construction of a so-called \emph{adjoint} system and depending on the whole trajectory of the HK model).

\subsection{Extensions and related models}

Recently a lot of alternative models based on the ideas of bounded confidence and extending the HK model in different ways, have been proposed. One extension,
known as the Deffuant-Weisbuch model, will be considered in Section~\ref{sec.gossip}, dealing with asynchronous gossip-based models. Focused on agent-based models, this tutorial also does not address statistical (Eulerian) bounded confidence models studied in~\cite{Blondel:2007,Blondel:2010,WedinHegarty:2015,LorenzSurvey:2007,Fortunato:2005,MirtaJiaBullo:2014,HendrickxOlshevsky:16}. Some other extensions are briefly summarized in this subsection.

\subsubsection{Continuous-time bounded confidence models}

Many results, available for the original HK model, have been extended to its \emph{continuous-time} counterparts. The direct extension of the scalar HK model~\eqref{eq.hk}, introduced in~\cite{Blondel:2010}, is as follows
\be\label{eq.hk-cont}
\dot x_i(t)=\sum_{j:|x_j(t)-x_i(t)|<d}(x_j(t)-x_i(t))\in\r
\ee
(following~\cite{Blondel:2009}, this model deals with open confidence intervals; modifications with closed intervals have been also considered~\cite{YangDimarogonas:2014}).

The differential equation~\eqref{eq.hk-cont} has a discontinuous right-hand side, which gives rise to the problem of \emph{solution existence}. As has been shown in~\cite{Blondel:2010}, the classical Carathe\'odory solution (with $x_i(t)$ absolutely continuous for $t\ge 0$ and, moreover, differentiable everywhere except for a countable set of points)
exists for \emph{almost all} initial conditions $x(0)$. Using the result of Lemma~\ref{lem.converge2}, it can be easily shown that every such solution converges as $t\to\infty$.
In general, the solutions does not reach consensus, however, consensus has been proved in the situation where the initial interaction graph $\g(x(0))$ is ``densely'' connected~\cite{YangDimarogonas:2014}. Alternatively, one may consider generalized solutions (replacing, as usual, the discontinuous right-hand side by a differential inclusion).
In~\cite{CeragioliFrasca:2012}, the existence of \emph{Krasovskii} solutions for any initial conditions and their convergence have been shown. Krasovskii solution is not uniquely determined by its initial condition and, in general, may exhibit some ``pathological'' behavior (e.g. the solution starting at an equilibrim point may leave it and converge to another equilibrium).
To avoid numerical instabilities, caused by the discontinuities, the following ``smoothed'' modification of the HK model can be introduced~\cite{CeragioliFrasca:2012,YangDimarogonas:2014}
\be\label{eq.hk-smooth}
\dot x_i(t)=\sum_js(x_j-x_i)(x_j(t)-x_i(t))\in\r,
\ee
where $s:\r\to \r_+$ stands for some even continuous function (similar multidimensional models, extending~\eqref{eq.jabin} to the case of continuous time, have been examined in~\cite{MotschTadmor:2013,JabinMotsch:2014,Stamoulas:15}).
Smoothed and discontinuous bounded confidence models inherit many properties of the original HK model, e.g. the order preservation property~\cite{CeragioliFrasca:2012}.
Similar to the discrete-time model~\cite{Blondel:2009}, criteria for the equilibria's local stability can be obtained~\cite{CeragioliFrasca:2012,Stamoulas:15}.
Unlike the discrete-time model, the models~\eqref{eq.hk-cont} and~\eqref{eq.hk-smooth} also preserve the ``average'' opinion $\bar x(t)=n^{-1}(x_1(t)+\ldots+x_n(t))$ since, obviously, $\dot{\bar x}(t)=0$ almost everywhere.

\subsubsection{Effects of stubborness}

As has been discussed in Part I of this tutorial~\cite{ProTempo:2017-1}, the dynamics of the French-DeGroot model changes dramatically in presence of stubborn individuals (keeping their opinions unchanged). Further relaxation of the stubborness concept leads to the Friedkin-Johnsen (FJ) model, where some agents can be ``partially'' stubborn (prejudiced). Such agents assimilate the others' opinions, being at the same time ``anchored'' at their initial opinions and factoring them into every step of the opinion iteration. Similar extensions have been suggested for the HK model.

In their work~\cite{HegselmannKrause:2006}, Hegselmann and Krause have proposed a model that inherits both the HK and FJ models. Consider $n$ agents with $m$-dimensional opinions $x_1,\ldots,x_n\in\r^m$ and fix one point $T\in\r^m$ in the opinion space referred to as the ``truth''. Assigning agent $i$ with a constant $\la_i\in [0,1]$ that characterizes the attractiveness of the truth for this agent and being a counterpart of the susceptibility in the FJ model~\cite{ProTempo:2017-1}, the dynamics from~\cite{HegselmannKrause:2006} is as follows
\be\label{eq.hk-truth}
x_i(t+1)=\frac{\la_i}{|I_i(x(t))|}\sum_{j\in I_i(x(t))}x_j+(1-\la_i)T,
\ee
where the $I_i(x)$ stand for the sets of trusted individuals~\eqref{eq.trust-multidim}. As discussed in the more recent work~\cite{HegselmannKrause:2015}, the ``truth'' value may be considered as
some external signal, influencing the system. The agents with $1-\la_i>0$ are referred in~\cite{HegselmannKrause:2006} as \emph{truth seekers}. In the absence of truth seekers, \eqref{eq.hk-truth}
boils down to the usual HK model~\eqref{eq.hk}. The truth seekers with $\la_i=T$ are \emph{stubborn}: $x_i(t)\equiv T$ for $t\ge 1$. Comparing the model~\eqref{eq.hk-truth} with the FJ model,
one notices two principal differences: first, the influence graph is distance-dependent (giving rise to the convergence problem\footnote{In general, opinions in the FJ model with time-varying influence graph can oscillate even when the graph remains strongly connected and some agents have $\la_i<1$, see~\cite{ProTempoCao16-2}.}) and, second, the agents have equal prejudices.

The most general result, concerned with the convergence of the model~\eqref{eq.hk-truth}, is as follows.
\begin{thm}\cite{Chazelle:11,HegselmannKrause:2015}\label{thm.hk-truth}
The opinions of all truth seekers ($\la_i<1$) converge to the truth\footnote{In the scalar case, this convergence was proved in~\cite{HegselmannKrause:2006,KurzRambau:2011}.} $x_i(t)\xrightarrow[t\to\infty]{} T$. The opinions of the remaining agents (with $\la_i=0$)
either converge to $T$ or \emph{stabilize} in finite time at some values $\bar x_i$ such that $|\bar x_i-T|\ge d$.
\end{thm}

Notice that convergence of the opinions to the truth value is usually asymptotical but not finite-time, as can be easily shown for the system of $n=2$ agent, one of them being stubborn
$x_1(t)\equiv T\in\r$ and the other starting at some point $x_2(0)\in (T-d,T+d)$. Some conservative estimates for the convergence rate have been obtained in~\cite{Chazelle:11}.
A natural question when the opinions reach consensus at $T$ in presence of agents with $\la_i=0$ still remains open. Numerical results, reported in~\cite{HegselmannKrause:2015} for the special case where all agents are either stubborn ($\la_i=1$) or do not seek the truth ($\la_i=0$) have revealed a highly non-trivial and counter-intuitive dependence between the number of stubborn agents, the confidence bound $d$ and the number of clusters. In particular, for some $d$ consensus is reached for small number of stubborn individuals and is destroyed as their number increases.

A more general class of models with stubborn individuals have been studied in~\cite{WangChazelle:17-TAC} by using Lyapunov techniques. In~\cite{WangChazelle:17-TAC}, the class of ``inertial'' HK models has been studied\footnote{Note that the ``inertial'' bounded confidence models have been also introduced and numerically studied in~\cite{UrbigLorenz:08,FuZhangLi:2015}.}, obeying the equations
\be\label{eq.hk-inert}
x_i(t+1)=(1-\la_i)x_i(t)+\frac{\la_i}{|I_i(x(t))|}\sum_{j\in I_i(x(t))}x_j.
\ee
Here $\la_i\in [0,1]$ (referred in~\cite{WangChazelle:17-TAC} to as the coefficient of \emph{inertia}); the agents with $\la_i(t)\equiv 0$ are stubborn (in~\cite{WangChazelle:17-TAC}, they are called ``close-minded''). For the general system~\eqref{eq.hk-inert}, extensions of Proposition~\eqref{prop.energy} and the inequality~\eqref{eq.k2} have been established in~\cite{WangChazelle:17-TAC}. These results allow to prove that the HK model with stubborn agents, that is, the system~\eqref{eq.hk-inert} where each agent has either $\la_i=0$ or $\la_i=1$, always converges.
\begin{thm}\cite{WangChazelle:17-TAC}\label{thm.hk-inert}
The opinions in the system~\eqref{eq.hk-inert} with $\la_i\in\{0,1\}$ asymptotically converge.
\end{thm}
Notice that, unlike Theorem~\ref{thm.hk-truth}, stubborn agents in the model~\eqref{eq.hk-inert} need not have identical opinions. In the case where $\la_i=1$ for any $i$, Theorem~\ref{thm.hk-inert}
implies Theorem~\ref{thm.hk-mult} (for the Euclidean norm). In~\cite{WangChazelle:17-TAC}, the result of Theorem~\ref{thm.hk-inert} has been extended to ``anchored'' HK systems, where opinions of each agent consists of a ``mobile'' part and static part; such systems appear to be equivalent, in some sense, to a special case of the heterogeneous model~\eqref{eq.jabin1}.

\subsubsection{Asymmetric interactions}

An important property of the HK model~\eqref{eq.hk},\eqref{eq.trust-multidim}, dramatically simplifying its analysis, is the \emph{symmetry} of interactions. The influence graph $\g(x(k))$ is undirected, that is, at each step $k$ every two agents $i$ and $j$ either mutually influence each other or are independent. The modifications of the HK models with \emph{asymmetric} interactions are much more complicated, and many of their properties observed in experiments are still waiting for mathematical investigation.

The simplest asymmetric bounded confidence model, proposed in~\cite{Krause:2002}, deals with scalar opinions $x_i\in\r$ and asymmetric confidence intervals, that is, the set of trusted individuals
is defined as follows
\be\label{eq.trust-asymm}
I_i(x)=\{j: -d_l\le x_j-x_i\le d_r\},\quad d_l,d_r>0.
\ee
Obviously, in the case where $\ve_l\ne\ve_r$ the graph $\g(x(k))$ can be \emph{directed} (which makes it impossible to apply Lemma~\ref{lem.converge1}, ensuring convergence). Also, the order of opinions in general is not preserved~\cite{BhatChazelle:2013,CoulsonTouri:2015}. Nevertheless, modification of the proof discussed in Subsect.~\ref{sec.krause}.1 allows to show that the model~\eqref{eq.trust-asymm} terminates in finite time; moreover, this holds for the more general class of asymmetric models with \emph{heterogeneous} confidence intervals as follows
\be\label{eq.trust-asymm+}
I_i(x)=\{j: -d+\eta_i\le x_j-x_i\le d\},\quad d,\eta_i>0.
\ee
\begin{thm}\cite{BhatChazelle:2013,CoulsonTouri:2015}\label{thm.hk-asymm1}
Assume that $\eta_i\ge 0\,\forall i$ and $\eta=\max_i\eta_i<d$. Then the asymmetric HK model~\eqref{eq.hk},\eqref{eq.trust-asymm+} terminates in finite time.
\end{thm}
As has been shown in~\cite{CoulsonTouri:2015}, the termination time can be estimated as $O(n^3)+O(n^2)\ln(1-\zeta)$, where $\zeta=\eta/d<1$, in particular, if
$\zeta<1-\exp(-O(n))$ then the convergence time of the model is $O(n^3)$ like in the symmetric HK model ($\eta_i=0$).

Obviously, Theorem~\ref{thm.hk-asymm1} remains valid for the confidence intervals $(x_i-\ve,x_i+\ve-\eta_i)$ (which can be proved by changing the signs of opinions $x_i\mapsto -x_i$).
However, allowing \emph{both} left and right endpoints of the confidence intervals to be heterogeneous
\be\label{eq.trust-asymm1}
I_i(x)=\{j: -d_l^i\le x_j-x_i\le d_r^i\},\quad d_l^i,d_r^i>0,
\ee
one arrives at a very complicated system, still waiting for thorough analysis. Most typically, the confidence intervals are symmetric
$d_l^i=d_r^i=d^i$ or, dealing with multidimensional opinions $x_i\in\r^m$, one has\footnote{Along with bounded confidence model, a ``bounded influence'' model can be considered~\cite{MirtaBullo:2012}, replacing~\eqref{eq.trust-multi1} by $I_i(x)=\{j: \|x_j-x_i\|\le d^j\}$; the two models are equivalent in the homogeneous case, being quite different when $d^i\ne d^j$.}
\be\label{eq.trust-multi1}
I_i(x)=\{j: \|x_j-x_i\|\le d^i\}.
\ee
Whereas the model~\eqref{eq.hk},\eqref{eq.trust-multi1} has been proposed simultaneously with its homogeneous counterpart ($d^i=d$)~\cite{Krause:2002},
its behavior in general ``remains a mystery''~\cite{WangChazelle:17-TAC}. Unlike the special case of the homogeneous model with stubborn agents ($d^i\in\{d,0\}\,\forall i$)
discussed in the previous subsection, in general the convergence has been proved only for special solutions~\cite{MirtaBullo:2012,EtesamiBasar:2015}, although simulations show that
the convergence is a generic property of the heterogeneous HK model~\cite{MirtaBullo:2012,Lorenz:2010} and its modifications~\cite{FuZhangLi:2015,ZhaoZhang:16,ChenGlass:16}. An interesting phenomenon reported in~\cite{Lorenz:2010} is emergence of consensus in the homogeneous HK model after injecting a very small proportion of agents with different confidence bound.
Sufficient conditions for consensus in some heterogeneous HK models have been proposed in~\cite{Vasca:17,Vasca:2016}.

\subsubsection{Other extensions}

The idea of bounded confidence, allowing to explain the phenomenon of persistent disagreement between opinions, has inspired numerous novel models of opinion formation. Most of them have been
studied numerically and their mathematical properties have not been fully understood. For this reasons, the relevant works are only briefly mentioned.

Obviously, in reality social actors do not know the exact values of the others' opinions, which gives rise to the problem of \emph{robustness} against various disturbances.
Numerical simulations show high sensitivity of the HK dynamics to inaccuracies in the floating point arithmetic~\cite{HegselmannKrause:2015}. This is consonant with the recent analytic result~\cite{SuChenHong:17}, showing that small additive noises destroy clusters in the HK model and lead to ``quasi-consensus''.
Similar effect is reported in~\cite{WangChazelle:17} for stochastic differential equations,
extending the continuous-time model~\eqref{eq.hk-cont}. The model's ability to generate disagreement is however regained, allowing some \emph{non-local} random interactions between the agents (an agent's opinion is not confined to the confidence interval)~\cite{Pineda:2013,Baccelli:2014,Nakamura:16,ZhangHong:17}. Some extensions of the ``truth-seeking'' model~\eqref{eq.hk-truth}, allowing random noises, have been proposed in~\cite{ZhaoZhang:16,DouvenRiegler:2010-1,DouvenRiegler:2010-2}.

Bounded confidence models appear to be related with \emph{community detection} algorithms in graphs~\cite{MorarescuGirard:2011}, Bayesian algorithms for distributed decision making~\cite{Varshney:2014} and algorithms of data clustering~\cite{Oliva:2015}. Bounded confidence models have been proposed for dynamics of ``uncertain'' opinions (standing for \emph{intervals}
of possible values)~\cite{LiangDongLi:16} and ``linguistic'' opinions representing words of a formal language~\cite{DouvenRiegler:2010-3,DongChen:16}.

The HK model belongs to a family of so-called \emph{influence systems}, introduced by B. Chazelle~\cite{Chazelle:2012,Chazelle:2015,Chazelle:2015-1} and generalizing a number of multi-agent algorithms arising in social and natural science. An influence system corresponds to the distributed protocol of iterative averaging, similar to
the French-DeGroot and Abelson's models and their nonlinear counterparts~\cite{Moro:05,LinFrancis:07,MatvPro:2013,Muenz:11}, over a \emph{state dependent} graph.
The existence of an arc in such a graph (that is, interaction between a pair of agents) is determined by some system of algebraic inequalities (strict or non-strict) with rational coefficients.
The fundamental property of influence systems with \emph{bidirectional} graphs is their asymptotic convergence (which has been shown for \emph{homogeneous} HK model), whereas influence systems with
\emph{directed} graphs can exhibit very complex dynamics, being e.g. chaotic or Turing-complete (able to simulate any Turing machine)~\cite{Chazelle:2015}.
However, these ``irregular'' behaviors appear to be non-robust against small random perturbations, making almost all the trajectories of an influence system asymptotically periodic~\cite{Chazelle:2015}.

It should be noticed that introduction of the distant-based influence weights is not the only way to describe effects of homophily and biased assimilation in social groups, as illustrated by the recent work~\cite{Dandekar:2013} that advocates a novel nonlinear extension of the DeGroot model to explain opinion polarization.

%% file: 4gossip.tex
\section{Randomized Gossip-based Models}\label{sec.gossip}

The models considered in the previous sections adopt an implicit assumption of \emph{synchronous} interactions among the agents. The agents simultaneously display their opinions to each other
and simultaneously update them. Evidently, even for small-group discussions this assumption is unrealistic; as noticed in~\cite{FriedkinJohnsen:1999}, ``interpersonal influences do not occur
in the simultaneous way... and there are more or less complex sequences of interpersonal influences in the group''. One approach to portray these asynchronous and interactions among social actors is known as \emph{gossiping}, assuming that agents interact not simultaneously but in \emph{in pairs}. At any step, two\footnote{Some results, discussed in this section, can be extended to the case of \emph{synchronous gossiping}~\cite{FagnaniZampieri:2008} where several dyadic interactions occur during each interaction session. For simplicity, we confine ourselves to the case of asynchronous interactions.} agents interact (e.g. meet each other at some public place or communicate via phone/e-mail), after which one or both of their opinions can be changed.

Interest to gossip protocols has been stirred up by the following \emph{gossiping} (or ``telephone'') problem in graph theory~\cite{Tijdeman:1971,HajnalGossip:1972,BakerShostak:1972}. Suppose that each of $n$ people knows an item  of scandal, which  is  not  known  to  any  of  the  others.  They  communicate  by  telephone,  and
whenever two individuals make  a call, they pass  on to each other, as much scandal as they know  at  that  time.  How  many  calls  are  needed  before  all  the  individuals know  all  the scandal? A more general problem with unidirectional communication has been addressed in~\cite{Harary1974Gossip-1,Harary1974Gossip-2}. In the case of bidirectional information exchange (requiring undirected communication topology) and $n\ge 4$ the worst-case number of calls is $2n-4$, whereas unidirectional communication over strongly connected directed graph requires, in general, $2n-2$ calls~\cite{Harary1974Gossip-1,Harary1974Gossip-2}. A survey of results on gossiping and a similar \emph{broadcasting} problem (an item of information, known by one agent, has to
 be transmitted to all other agents) can be found in~\cite{GossipReview88}.

The pairwise gossiping  interactions between the agents need not be random; consensus and other problems of multi-agent control can be solved e.g. by using \emph{periodic} gossiping~\cite{LiuMouMorse:11} and other distributed algorithms with deterministic \emph{asynchronous} events~\cite{CaoMorse:08Part2,CaoMorse:08-1}. In this tutorial, we focus on
\emph{randomized} gossip-based models of opinion formation, where random choice of the agents interacting at each step mimics spontaneity of real social interactions.

In this section, we suppose that the reader is familiar with the basic concepts of probability theory (probability spaces, random variables and their distributions, expectation and moments, convergence in probability and almost surely etc.)~\cite{ShiriaevBook}. Henceforth $\P(A)$ denotes the probability of an event $A$ and $\E f$ stands for expectation of a random variable $f$.

\subsection{Gossip-based consensus}

In spite of relatively slow convergence, gossip-based consensus algorithms have attracted a lot of attention, being simple and very parsimonious in use of communication resources.
The simplest linear gossiping algorithms can be considered as special cases of the French-DeGroot model~\eqref{eq.degroot1}
with \emph{random} i.i.d.\footnote{\emph{Independent and identically distributed}} stochastic matrices $W(k)$. We first discuss some properties of such a randomized dynamics.
\begin{defn}
In the system~\eqref{eq.degroot1}, opinions are said to synchronize in probability, almost surely or in the $p$-th moment ($p>0$) if for any $i,j=1,\ldots,n$ and any (deterministic) initial condition the sequence $x_i(k)-x_j(k)$ converges in the corresponding sense, i.e.
\[
\begin{gathered}
\P(|x_i(k)-x_j(k)|\ge\ve)\xrightarrow[k\to\infty]{} 0\,\forall\ve>0\;\;\text{(in probability)}\\
\P\left(\lim_{k\to\infty}|x_i(k)-x_j(k)|=0\right)=1\quad\text{(almost surely)}\\
\E|x_i(k)-x_j(k)|^p\xrightarrow[k\to\infty]{} 0\quad\text{(in the $p$-th moment)}
\end{gathered}
\]
\end{defn}

The following fundamental result~\cite{SalehiJadbabaie:2008,FagnaniZampieri:2008,SongChenHo:11} (see also special cases in~\cite{HatanoMesbahi:05,Wu:06})
extends Proposition~\ref{prop.ergo} to the case of randomized French-DeGroot model, reducing
it to a \emph{deterministic} model.
\begin{lem}\label{lem.random}
Consider the system~\eqref{eq.degroot1} with i.i.d. stochastic matrices $W(k)$ and denote $\bar W=\E W(0)$. Then the following statements are equivalent
\begin{enumerate}[a)]
\item the deterministic French-DeGroot model
\be\label{eq.degroot-compari}
x(k+1)=\bar Wx(k),\quad k\ge 0
\ee
reaches consensus, i.e. the stochastic matrix $\bar W$ is fully regular (or stochastic indecomposable aperiodic, SIA)~\cite{ProTempo:2017-1};
\item the opinions reach consensus almost surely
\be\label{eq.conse-random}
\P(\exists c\in\r:\lim_{k\to\infty}x(k)=c\mathbbm{1}_n)=1\quad\forall x(0);
\ee
\item the opinions synchronize almost surely;
\item the opinions synchronize in probability;
\item the opinions synchronize in the $p$-th moment for some $p\in [1,\infty)$.
\end{enumerate}
Furthermore, if (a)-(e) hold and $W(k)$ have a common nonnegative left eigenvector $w^{\top}$, such that $w^{\top}\mathbbm{1}_n=1$, then in~\eqref{eq.conse-random} one has $c=w^{\top}x(0)$.
\end{lem}

The equivalence between (a) and (c)-(e) has been proved in~\cite{SongChenHo:11}. Obviously, (b) implies (c). The last statement of Lemma~\ref{lem.random} and the
implication (a)$\Longrightarrow$(b) has been established
in~\cite{SalehiJadbabaie:2008} under the additional assumption that $W(k)$ have strictly positive diagonal; the latter assumption can also be discarded~\cite{FagnaniZampieri:2008}
(formally,~\cite{FagnaniZampieri:2008} establishes the implication (a)$\Longrightarrow$(b) only for the case of irreducible $\bar W$, but the reducible case can be considered similarly).
Consensus criteria, similar in spirit to Lemma~\ref{lem.random}, have been also established for non-stationary randomized models (where the i.i.d. assumption fails)~\cite{SalehiJadbabaie:2010,TouriNedic:11}.

\begin{rem}\label{rem.random}
Under the assumptions of Lemma~\ref{lem.random}, $W(k)$ is independent of $x(k)$ (which depends only on $W(1),\ldots,W(k-1)$), which implies that the vectors $\E x(k)$  obey the deterministic model~\eqref{eq.degroot-compari}
\[
\E x(k+1)=\E W(k)\E x(k)=\bar W\E x(k)\quad\forall k,
\]
in particular, the opinions' expectations $\E x_i(k)$ converge to consensus if (a)-(e) hold.
\end{rem}

Consider now the following asynchronous gossip algorithm~\cite{FagnaniZampieri:2008}. Consider a group of $n$ agents with opinions $x_1(k),\ldots,x_n(k)\in\r$ and a stochastic matrix $P=(p_{ij})$ with zero diagonal $p_{ii}=0$. At each step $k$ of the opinion iteration, one agent $i=i(k)$ is randomly activated; we assume that the sequence $i(k)$ is i.i.d. and uniformly distributed in $\{1,\ldots,n\}$. With probability $p_{ij}$, the active agent $i$ interacts with agent $j$ and updates its opinion as follows
\be\label{eq.degroot-goss}
x_i(k+1)=(1-\gamma_i)x_i(k)+\gamma_{i} x_j(k),
\ee
where $\gamma_{i}\in(0,1)$ is a constant, describing the ``trust'' of individual $i$ in his/her neighbors. The opinions of the other agents (including $j$) remain unchanged
\be\label{eq.degroot-goss1}
x_l(k+1)=x_l(k)\quad\forall l\ne i(k).
\ee

It can be easily shown that the algorithm~\eqref{eq.degroot-goss},\eqref{eq.degroot-goss1} is a special case of~\eqref{eq.degroot1}, where $W(k)$ are i.i.d. stochastic matrices, attaining
one of the values $W^{ij}=I_n+\gamma_i(e_j-e_i)e_i^{\top}$ with the probability and $\P(W(k)=W^{ij})=n^{-1}p_{ij}$ (here $e_i\in\r^n$ is the column vector with all components equal to $0$ except
for the $i$th component, which is equal to $1$). Introducing the diagonal matrix $\Gamma=\diag(\gamma_1,\ldots,\gamma_n)$, we have
\[
\bar W=\E W(1)=I_n-n^{-1}\Gamma+n^{-1}\Gamma P.
\]
Obviously, the diagonal entries of $\bar W$ are positive and when $i\ne j$, one has $\bar w_{ij}>0\Leftrightarrow p_{ij}>0$. Hence, the system~\eqref{eq.degroot-compari} reaches consensus if and only if the graph $\g(P)$ has a directed spanning tree (see~\cite{ProTempo:2017-1}). Lemma~\ref{lem.random} yields in the following consensus criterion.
\begin{cor}\label{cor.gossip}
For any gains $\gamma_1,\ldots,\gamma_n$ and stochastic matrix $P$, such that the graph $\g(P)$ has a directed spanning tree, the protocol~\eqref{eq.degroot-goss},\eqref{eq.degroot-goss1} provides
consensus of opinions with probability $1$~\eqref{eq.conse-random}.
\end{cor}

Along with unidirectional gossip algorithm, one may consider \emph{bidirectional} protocol where \emph{both} agents $i$,$j$ update their opinions. The simplest algorithm of this type has been examined in~\cite{Boyd:06}
\be\label{eq.boyd}
\begin{gathered}
x_i(k+1)=x_j(k+1)=\frac{x_i(k)+x_j(k)}{2}\\
x_l(k+1)=x_l(k)\quad\forall l\ne i,j.
\end{gathered}
\ee
It can be shown~\cite{Boyd:06} that in this situation one has
\[
\bar W=I-\frac{1}{2n}D+\frac{1}{2n}\left(P+P^{\top}\right),
\]
where $D$ is the diagonal matrix with entries $d_{ii}=1+\sum_{j}p_{ji}\le 1+n$. Consensus in~\eqref{eq.degroot-compari} is established if and only if the undirected graph, corresponding to $P+P^{\top}$, is connected.

A more advanced analysis of gossip-based consensus algorithms and overview of their applications are available in~\cite{AysalYildizScaglione:09,DimakisScaglione:10,FagnaniZampieri:2008,Boyd:06,LiuAndersonCaoMorse:13,LiuMouMorse:11} and references therein.

\subsection{Gossiping with stubborn agents}

As has been in discussed in Part I~\cite{ProTempo:2017-1}, the presence of several \emph{stubborn} agents in the static French-DeGroot model~\eqref{eq.degroot} yields in more interesting dynamics than conventional consensus algorithms exhibit: opinions do not reach consensus and typically split into several clusters (this holds when the matrix $W$ is \emph{regular}, i.e. the limit $\lim\limits_{k\to\infty}W^k$ exists). A natural question arises whether its \emph{gossip-based} counterpart exhibits a similar behavior.
The answer to this question appears to be \emph{negative}: as has been shown in~\cite{Acemoglu:2013}, in presence of stubborn agents the gossip algorithms
fail to provide almost sure convergence of the opinions, which keep on fluctuating in an ergodic fashion. At the same time, the distribution of the opinion vector $x(k)$ converges to some probability measure. In this subsection, we discuss similar results\footnote{Technically the model in~\cite{Acemoglu:2013} differs from the models considered in this subsection, e.g. it deals with dynamics on a continuous time scale, where interactions between pairs of connected agents are activated by clocks, each ticking at the times of an independent Poisson
process of certain rate. Some results from~\cite{Acemoglu:2013}, concerned with the characteristics of the stationary distribution, still have not been extended to the FJ model.}, concerning the asynchronous gossip-based version of the  Friedkin-Johnsen (FJ) model~\cite{FrascaTempo:2013,FrascaTempo:2015,FrascaIshiiTempo:2015,Parsegov2015CDC,Parsegov2017TAC}.

\subsubsection{Example: fluctuation between two stubborn leaders and Bernoulli convolution series}

We start with an example from~\cite{Acemoglu:2013}, demonstrating that in presence of stubborn agents the gossip algorithm does not provide convergence of the opinions. Consider two stubborn individuals with fixed opinions $x_1\equiv 0$ and $x_2\equiv 1$ and ``regular'' agent that can interact with both of them. At each step $k$, the regular agent chooses one of the stubborn neighbors $i=i(k)\in\{1,2\}$ with probability $1/2$ and shifts its own opinion towards the opinion of this neighbor
\[
x_3(k+1)=(1-\gamma)x_3(k)+\gamma x_{i(k)},\quad \gamma\in (0,1).
\]
Since $x_3(k)=(1-\gamma)^kx_3(0)+\gamma\sum_{s=0}^{k-1}(1-\gamma)^{s}x_{i(k-s)}$, it can be shown that the distribution of $x_3$ converges\footnote{Henceforth, by convergence of distributions we mean the standard \emph{weak} convergence: a sequence of probability measures $\P_k$ on the same $\sigma$-algebra converges to a measure $\P$, if $\E_{\P_k}f\to \E_{\P}f$ as $k\to\infty$ for any bounded random variable $f$.} as $k\to\infty$ to the distribution of
the random variable
\[
\bar x_3=\frac{\gamma}{1-\gamma}\bar\xi,\quad\bar\xi=\sum_{s\ge 1}(1-\gamma)^s\xi_s,
\]
where $\xi_j$ are i.i.d. Bernoulli random variables with probability $1/2$. The random variable $\bar\xi$ in the right-hand side is referred to as the \emph{Bernoulli convolution}~\cite{PeresSolomyak2000}. For $\gamma=1/2$, $\bar\xi$ is uniformly distributed on $[0,1]$. For almost all $\gamma\in (0,1/2)$ its distribution on [0,1] is absolutely continuous
(has a density), whereas for $\gamma>1/2$ it is supported on the Cantor set~\cite{PeresSolomyak2000}.

This example demonstrates, in particular, that consensus in Lemma~\ref{lem.random} cannot be replaced by \emph{convergence} of the opinions:
such a convergence in the deterministic model~\eqref{eq.degroot-compari}, in general, does not imply the convergence of the model~\eqref{eq.degroot1} with random $W(k)$.
At the same time, Remark~\ref{lem.random} implies that \emph{expected} values of the opinions converge. For the special gossip-based model considered below, the convergence of \emph{time averages}
can also be proved.

\subsubsection{The asynchronous gossip-based FJ model}

Recall that the FJ model~\cite{FriedkinJohnsen:1999} is characterized by two matrices $\La,W$, where $\La$ is a diagonal matrix of agents' susceptibilities to social influence, $0\le \La\le I_n$, and $W$ is a stochastic matrix of influence weights. The opinion vector $x(k)\in\r^n$ evolves as follows
\be\label{eq.fj}
x(k+1)=\La Wx(k)+(I-\La)u,\quad u=x(0).
\ee
The relations between the FJ model and the French-DeGroot model has been discussed in Part I~\cite{ProTempo:2017-1}, as well as the graphical conditions for the model's stability and convergent. In this subsection, we assume that the model~\eqref{eq.fj} is a stable\footnote{As shown in~\cite{Parsegov2017TAC,ProTempo:2017-1}, unstable FJ model (with $\rho(\La W)=1$) contains a subgroup of agents obeying the usual French-DeGroot model and independent of the remaining agents. Such situation is impossible, e.g. when $\La\ne I$ and $W$ is irreducible.} as a system with static input $u$, that is, $\La W$ is a Schur stable matrix $\rho(\La W)<1$, and hence the opinions in~\eqref{eq.fj} converge
\be\label{eq.fj-final}
x(k)\xrightarrow[k\to\infty]{} \bar x=(I-\La W)^{-1}(I-\La)u.
\ee

Numerous experiments with small and medium-size group~\cite{FriedkinJohnsen:1999,Friedkin:2012,FriedkinJohnsenBook,FriedkinBullo:17} have confirmed the predictive power
of the FJ model, in particular,~\eqref{eq.fj-final} gives a good approximation for the real distribution of final opinions.
This means that the asynchronous gossip-based counterpart of the FJ model should also provide (in some sense) the correspondence~\eqref{eq.fj-final} between the ``prejudice'' vector $u=x(0)$ and the outcome of the opinion formation process. We consider one such gossip-based model, proposed in~\cite{Parsegov2017TAC} and generalizing the protocols from~\cite{FrascaTempo:2013,FrascaTempo:2015,FrascaIshiiTempo:2015}.

Let $\g[W]=(\mathcal V,\mathcal E)$ stand for the interaction graph of the network and consider two matrices
\be\label{eq.gammas}
\Gamma^1=\Lambda W,\quad \Gamma^2=(I-\Lambda)W.
\ee
Consider the following asynchronous gossip-based algorithm, similar in structure to~\eqref{eq.degroot-goss},\eqref{eq.degroot-goss1}.
At each step $k$, an arc is uniformly sampled in the set $\mathcal E$. If this arc is $(i,j)$, then agent $i$ communicates to agent $j$ and updates its opinion as follows
\be\label{eq.gossipC}
x_i(k+1)=(1-\gamma_{ij}^1-\gamma_{ij}^2)x_i(k)+\gamma_{ij}^1x_j(k)+\gamma_{ij}^2u_i.
\ee
Hence during each interaction the agent's opinion is averaged with its own prejudice and the neighbors' opinion; note, however, that the weight of the prejudice depends on both $\la_{ii}$ and $w_{ij}$. The other opinions (including opinion of agent $j$) remain unchanged
\be\label{eq.gossip2}
x_l(k+1)=x_l(k)\quad\forall l\ne i.
\ee

The following theorem shows that the gossiping opinion formation process~\eqref{eq.gossipC}, \eqref{eq.gossip2} inherits the properties of the FJ model \emph{on average}.
Let $\bar x(k)$ stand for the \emph{Ces\'aro} (or \emph{Polyak}) average
\be\label{eq.cesaro1}
\bar x(k):=\frac{1}{k+1}\sum_{l=0}^kx(l).
\ee

\begin{thm}\cite{Parsegov2017TAC}\label{thm.gossip1}
Let $\rho(\Lambda W)<1$ and $\Gamma^1,\Gamma^2$ be matrices from~\eqref{eq.gammas}. Then, there is the following correspondence between the
 gossip-based model~\eqref{eq.gossipC}, \eqref{eq.gossip2} and the deterministic model~\eqref{eq.fj}:
\begin{enumerate}
\item the steady expected value $\lim\limits_{k\to\infty}\E x(k)=\bar x$ coincides with~\eqref{eq.fj-final} for any $x(0)$;
\item the random process $x(k)$ is \emph{almost sure ergodic}
\[
\P(\lim\limits_{k\to\infty}\bar x(k)=\bar x)=1;
\]
\item for any $p>0$, the process $x(k)$ is $L^p$-ergodic
\[
\E|\bar x(k)-\bar x|^p\xrightarrow[k\to\infty]{} 0.
\]
\end{enumerate}
Additionally, the process $x(k)$ converges in distribution to some random vector $x_{\infty}$ with $\E x_{\infty}=\bar x$, whose distribution is determined by $\La,W$.
\end{thm}

As noticed in~\cite{Parsegov2017TAC}, the statements of Theorem~\ref{thm.gossip1} remain valid for any matrix $\Gamma^2$ such that $0\le \gamma_{ij}^2\le 1-\gamma_{ij}^1$, $\sum_{j=1}^n\gamma_{ij}^2=1-\la_{ii}$ and $\gamma_{ij}^2=0$ as $(i,j)\not\in\mathcal E$. In~\cite{Parsegov2017TAC}, multidimensional gossip-based models have been also considered,
where opinions represent individuals' positions on several interrelated topics (belief systems~\cite{FriedkinPro2016}).
In~\cite{FrascaTempo:2013}, the estimate for the convergence rate (and variance) of $\bar x(k)$ has been obtained
\[
\E|\bar x(k)-\bar x|^2\le\frac{\chi}{k+1},
\]
where $\chi$ depends on $u$ and the spectral radius $\rho(\La W)$ (in fact, $\chi\to\infty$ as $\rho\to 1$). The proof of Theorem~\ref{thm.gossip1} is based on the elegant
results of~\cite{FrascaTempo:2015}, dealing with the properties of randomized affine system
\be\label{eq.frasca}
\xi(k+1)=M(k)\xi(k)+u(k),
\ee
with special i.i.d. sequences of the matrices $M(k)$ (non-stochastic) and vectors $u(k)$ with finite expectations.
The key property of the system~\eqref{eq.frasca},  needed for the existence of a stationary distribution and ergodicity of its solutions is the decomposability
\[
\E M(k)=\alpha I+(1-\alpha)P,
\]
where $\alpha\in (0,1)$ and $\det (I-P)\ne 0$.

\begin{figure}[h!]
\centering
\begin{subfigure}[t]{0.75\columnwidth}
        \centering
        \includegraphics[width=\columnwidth]{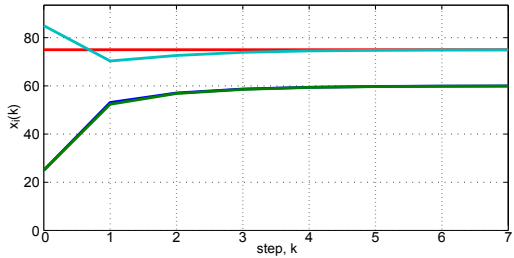}
        \caption{Deterministic model}
    \end{subfigure}\hfill
    \begin{subfigure}[t]{0.75\columnwidth}
        \centering
        \includegraphics[width=\columnwidth]{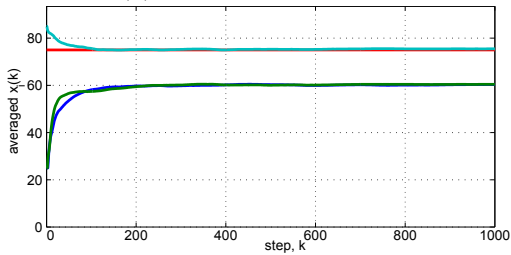}
        \caption{Averaged random opinions $\bar x(k)$}
    \end{subfigure}
\begin{subfigure}[t]{0.75\columnwidth}
        \centering
        \includegraphics[width=\columnwidth]{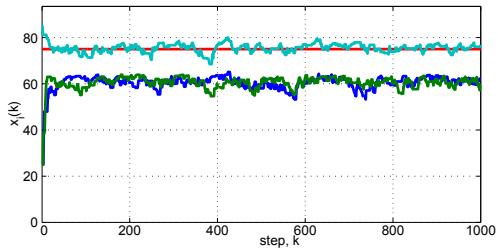}
        \caption{Random opinions $x(k)$}
    \end{subfigure}\hfill
\caption{Deterministic FJ model~\eqref{eq.fj} vs. the asynchronous gossip-based model~\eqref{eq.gossipC}, \eqref{eq.gossip2}.}\label{fig.fj-gossip}
\end{figure}

Similar to the example with three agents from previous subsection, opinion vectors $x(k)$ usually do not converge and keep on fluctuating, as demonstrated by the following example~\cite{Parsegov2017TAC}. Consider $n=4$ agents with the initial opinions $u=x(0)=(25,25,75,85)^{\top}$ and the matrix $W$, observed in a real social group~\cite{FriedkinJohnsen:1999}
\[
  W = \begin{bmatrix}
  0.220 & 0.120 & 0.360 & 0.300 \\
   0.147 & 0.215 & 0.344 & 0.294 \\
   0 & 0 & 1 & 0 \\
   0.090 & 0.178 & 0.446 & 0.286
\end{bmatrix}.
\]
The matrix $\La$ is defined by the ``coupling condition'' $\La=I-\diag W$. The final opinion vector~\eqref{eq.fj-final} of the FJ model is $\bar x\approx (60,60,75,75)^{\top}$, i.e.
agents $1$ and $2$ form their own cluster, whereas stubborn agent 3 attracts the opinion of agent 4. Fig.~\ref{fig.fj-gossip} illustrates the trajectories of the deterministic and randomized FJ models, as well as the Ces\'aro averages $\bar x(k)$. As shown in Fig.~\ref{fig.fj-gossip}c, the opinions fluctuate (except for the opinion of stubborn agent 3).

A similar behavior is demonstrated by the randomized algorithms for PageRank computation~\cite{IshiiTempo:2010,IshiiTempo:2014,YouTempoQui:2017}, whose relation to
the asynchronous gossip-based FJ model has been disclosed in~\cite{FrascaIshiiTempo:2015,FrascaTempo:2015}.

\subsection{The Deffuant-Weisbuch model}

Along with the HK model discussed Section~4, the \emph{Deffuant-Weisbuch} (DW) model is among the most illustrious models of opinion formation.
Being a gossip-based counterpart of the HK model, the DW model in fact was proposed in~\cite{DeffuantWeisbuch:2000} independent of
Krause's work~\cite{Krause}, employing the same principle of bounded confidence. In spite of many numerical results and experimental observations, dealing with the behavior of the DW model and its modifications over complex networks~\cite{DeffuantWeisbuch:2000,DefuantWeisbuch:2002,Defuant:2006,Weisbuch:2004,WeisbuchDeffuant:2005,Kurmyshev:2011,Kurmyshev:17},
the compound of randomness and nonlinear dynamics makes these models very hard for mathematical investigation, and analytic results explaining their behavior are limited.

The original DW model has a structure similar to the symmetric gossiping protocol~\eqref{eq.boyd}, however, the interaction graph is coincident with the graph $\g(x(k))$
introduced for the HK model in Section~4, being \emph{state-dependent} and determined by the distances between the opinions.
Similar to the HK model, the DW model~\cite{DeffuantWeisbuch:2000} deals with a group of $n$ agents,
each having the \emph{confidence bound} $d>0$ (referred also to as the agent's ``threshold''~\cite{DeffuantWeisbuch:2000} or ``uncertainty''~\cite{DefuantWeisbuch:2002}).
The opinions\footnote{For simplicity, we confine ourselves to scalar DW model, although vector opinions have been also introduced in~\cite{DeffuantWeisbuch:2000}.} $x_i\in\r$ are updated in accordance with the gossiping procedure as follows.

At each step, a pair of agents $(i,j)$ is chosen randomly. These two agents interact if and only if their opinions are close.
Denoting the indicator function of an event $\Omega$ with $\I({\Omega})$, the DW model is as follows
\be\label{eq.dw}
\begin{gathered}
x_i(k+1)=x_i(k)+\mu(x_j(k)-x_i(k))\I_k,\\
x_j(k+1)=x_j(k)+\mu(x_i(k)-x_j(k))\I_k,\\
x_l(k+1)=x_l(k)\quad\forall l\ne i,j,\\
\I(k)=\I(|x_j(k)-x_i(k)|\le d).
\end{gathered}
\ee
The constant $\mu\in (0,1)$ is called \emph{convergence parameter}~\cite{DeffuantWeisbuch:2000} and stands for the attraction between the opinions. Similarly to the models from the previous subsections,~\eqref{eq.dw} can be considered as the French-DeGroot model with \emph{state-dependent} random matrices $W(k)=W(x(k))$. Such matrices are no longer i.i.d. since $W(k)$ depends on $W(k-1)$, however, $x(k)$ remains a Markov process, and for any its realization the matrices $W(k)$, as can be easily shown, satisfy the conditions of Lemma~\ref{lem.converge1}, entailing in the following.
\begin{lem}\label{cor.dw}
The opinions in the model~\eqref{eq.dw} converge almost surely, that is
\[
\P(\exists \bar x_i=\lim_{k\to\infty}x_i(k))=1\quad\forall i.
\]
Furthermore, for any $i,j$ one almost surely has either $|\bar x_i-\bar x_j|\ge d$ or $\bar x_i=\bar x_j$.
\end{lem}

The first statement of Lemma~\ref{cor.dw} is immediate from Lemma~\ref{lem.converge1}. To prove the second statement, consider a pair of indices
 $(i_0,j_0)$. Due to the Borel-Cantelli lemma~\cite{ShiriaevBook}, $(i,j)=(i_0,j_0)$ infinitely often with probability 1. Hence either agents $(i_0,j_0)$ are infinitely connected in the time-varying
 graph $\g(W(k))$ (and then $\bar x_{i_0}=\bar x_{j_0}$) or the sequence $k_s\to\infty$ exists such that $|x_{i_0}(k_s)-x_{j_0}(k_s)|>d$, i.e. $|\bar x_{i_0}-\bar x_{j_0}|\ge d$.

An important and non-trivial result from~\cite{ZhangHong:13}, based on techniques from~\cite{TouriNedic:11}, shows that Lemma~\ref{cor.dw} retains its validity for the extensions of~\eqref{eq.dw}
that allow \emph{asymmetric} interactions among the agents. The simplest of such models is similar to~\eqref{eq.dw} yet assumes that only agent $i$ updates its opinion, and the other opinions remain constant
\be\label{eq.dw-asymm}
\begin{gathered}
x_i(k+1)=x_i(k)+\mu(x_j(k)-x_i(k))\I_k,\\
x_l(k+1)=x_l(k)\quad\forall l\ne i\\
\I_k=\I(|x_j(k)-x_i(k)|\le d).
\end{gathered}
\ee
More complicated are ``multi-choice''~\cite{ZhangHong:13} extensions of~\eqref{eq.dw-asymm}, allowing an ``active'' agent $i$ to interact with several neighbors. Lyapunov techniques allow to obtain some explicit estimates for the convergence rate of~\eqref{eq.dw} and its asymmetric counterparts~\cite{ZhangChen:2015}.

Similar to the case of HK model, \emph{heterogeneous} confidence bounds $d_i$ in the DW model lead to serious complications. The heterogeneous counterpart of~\eqref{eq.dw}, studied in~\cite{Lorenz:2010,MathiasHuetDeffuant:16}, is as follows
\be\label{eq.dw-heter}
\begin{gathered}
x_i(k+1)=x_i(k)+\mu(x_j(k)-x_i(k))\I_k^i,\\
x_j(k+1)=x_j(k)+\mu(x_i(k)-x_j(k))\I_k^j,\\
x_l(k+1)=x_l(k)\quad\forall l\ne i,j,\\
\I_k^i=\I(|x_j(k)-x_i(k)|\le d_i),\\
\I_k^j=\I(|x_j(k)-x_i(k)|\le d_j).
\end{gathered}
\ee
As reported in~\cite{MathiasHuetDeffuant:16}, in presence of ``extremists'' with very narrow confidence intervals, the model~\eqref{eq.dw-heter} exhibits \emph{ergodic} fluctuations of the opinions,
similar to the behavior of asynchronous gossip-based models with stubborn agents~\cite{Acemoglu:2013} and the FJ model~\eqref{eq.gossipC}.

Even more sophisticated are extensions of the DW models, involving nonlinear couplings among interacting agents (e.g. the ``relative agreement'' models~\cite{DefuantWeisbuch:2002,AmblardDeffuant:2004,Kurmyshev:2011,Kurmyshev:17}), dynamic or random confidence bounds~\cite{WeisbuchDeffuant:2002,Defuant:2006,Shang:2014}, noises~\cite{CarroToral:13,Baccelli:2017} and ``long-range'' interactions (agents can assimilate opinions beyond their confidence intervals)~\cite{ZhangHong:13,Baccelli:2017}. These models are beyond the scope of this tutorial, as well as the gossip-based counterparts of the ``truth-seeking'' model~\eqref{eq.hk-truth}, introduced in~\cite{Malarz:2006,PinedaBuendia:2015}.

%% file: 5negative.tex
\section{Disagreement via Negative Influence}\label{sec.negative}

The models discussed in the previous section extend the basic French-DeGroot and Abelson models, inheriting however the key idea of \emph{iterative averaging}.
Even though agreement is not always possible (due to the effects of stubborness, homophily etc.), the agents \emph{cooperate} in order to reach it,
always changing their opinions towards each other. In many systems, arising in economics, natural sciences and robotics such \emph{positive} (attractive) couplings among the agents coexist with \emph{negative} (repulsive) couplings, leading to the agents' distancing. These negative ties among the agents can correspond to their competition; multi-agent networks where agents can both cooperate and compete are sometimes called ``coopetitive''~\cite{Pache:13,HuZheng:2014}. Repulsive interactions lie in the heart of many biological system~\cite{CoyteScience:15,Samaniego:2017} and are vital
for collision avoidance in swarms and robotic formations~\cite{Romanczuk:12,YuChenCaoLuZhang:13,Asya:15}. Unlike the well-developed theory of \emph{cooperative} networks,
the studies on ``coopetitive'' networks are taking their first steps.

As a possible reason for disagreement of opinions, Abelson~\cite{Abelson:1967} mentions the \emph{boomerang effect}~\citep{HovlandBook,Allahverdyan:2014}, being an unintended consequence
of the persuasion process. An attempt to persuade a person sometimes shifts his/her opinion \emph{away} from the persuader's opinion. The boomerang effects can be partly explained by
personal insults~\citep{AbelsonMiller:1967} occuring during the discussion or the discussants' reactance\footnote{The reactance is an individual's ability to resist the persuasion, being heavily pressed to accept some attitude. This resistance may lead, under some circumstances, to adoption of an attitude, which is \emph{opposite} to the persuader's one.}~\citep{Dillard:2005}.
The theories of balance~\cite{Heider:1946} and cognitive dissonance~\citep{FestingerBook} explain the presence of negative influences by coevolution of the opinions and personal relations between individuals. An agreement with negatively evaluated person creates a psychological tension, and individuals can resolve
such tensions by ``moving their opinions away from a disliked source''~\cite{TakacsFlacheMas:16}. Models proposed in~\cite{FlacheMas:2008,Flache:2011,Feliciani:17} postulate
that negative ties between individuals arise due to large discrepances of their opinions. Unlike the bounded confidence models, an individual does not reject too dissimilar opinions but
rather shifts his/her opinion away from them. These and many other models of opinion formation under negative influence, offered in the literature and still awaiting rigorous analysis~\citep{Macy:2003,Mark:2003,Baldassarri:2007,Salzarulo:2006}, illustrate
that antagonism among the agents may lead to the community cleavage.

The ubiquity of such negative ties has not secured by experimental evidence, e.g. the recent experiments reported in~\cite{TakacsFlacheMas:16} does not support the aforementioned hypotheses, explaining negative influence to dissimilarities and disliking. Nevertheless, from the authors' point of view, models of opinion formation with antagonistic interactions deserve attention since they exhibit non-trivial behaviors and can lead to the development of novel mathematical theories. In this section, we consider the model proposed by Altafini~\cite{Altafini:2012,Altafini:2013} and its extensions. The idea of opinion polarization via structural balance has triggered the extensive research on ``bipartite'' consensus, synchronization and flocking~\cite{Valcher:14,FanZhangWang:14,Zhai:16}, extending the relevant results of multi-agent cooperative control to networks with antagonistic interactions.

\subsection{Balance theory}

The idea of Altafini's model has been inspired by the theory of \emph{structural balance}, pioneered in~\citep{Heider:1944,Heider:1946,CartwrightHarary:1956} and postulating the mutual dependence
between the interpersonal relations and the opinion formation: ``an attitude towards an event can alter the attitude towards the person who caused the event, and, if the attitudes
towards a person and an event are similar, the event is easily ascribed to the person''~\cite{Heider:1946}. Positive and negative evaluations of the individuals by each other are naturally
represented by \emph{signed graphs}.

\begin{defn} A signed graph is a triple $\g=(V,E,A)$, where $(V,E)$ is a graph and $A=(a_{ij})_{i,j\in V}$ is matrix, whose entries \emph{may be negative}, such that $a_{ij}\ne 0$ if and only if $(j,i)\in E$. An arbitrary matrix $A=(a_{ij})_{i,j\in V}$ corresponds to the unique signed graph $\g[A]=(V,E[A],A)$, where $E[A]=\{(i,j):a_{ji}\ne 0\}$.
\end{defn}

One can interpret positive arcs as relations of friendship or trust, negative ones standing for enmity or suspicion.
We call arc $(i,j)$ of a signed graph positive (negative) if its ``weight'' $a_{ij}$ is positive (negative). Henceforth, unless stated otherwise, we assume that all signed graphs
are \emph{sign-symmetric}~\cite{Altafini:2013} in the sense that $a_{ij}a_{ji}\ge 0$ (in other words, the relations of friendship and enmity are symmetric). This condition is necessary for
the structural balance and allows to simplify some constructions.

\begin{figure}[h]
\center
\includegraphics[width=0.75\columnwidth]{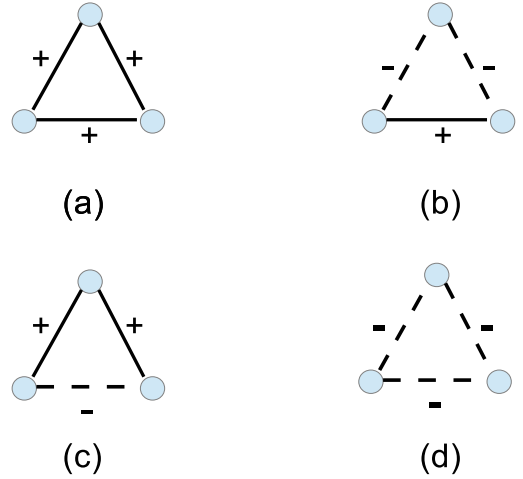}
\caption{Balanced vs. imbalanced triads}
\label{fig.triad}
\end{figure}

To illustrate the notion of structural balance, consider for the moment a \emph{complete} signed graph. The structural balance in such a graph can be defined as balance in each \emph{triad} (subgraph with three nodes), which is understood as follows~\cite{Heider:1946,CartwrightHarary:1956}.
\begin{defn}\label{def.triad}
A complete sign-symmetric graph is structurally balanced if each triad with nodes
$(i,j,k)$ is balanced in the sense that $a_{ij}a_{jk}a_{ki}>0$ (Fig~\ref{fig.triad}).
\end{defn}

A balanced triad (Fig.~\ref{fig.triad}a,b) represent either a triple of ``friends'' (a) or an alliance of two ``friends'' against a common ``enemy'' (b). The imbalance in a triad (Fig.~\ref{fig.triad}c,d), violating the ancient proverb \emph{the friend of my enemy is my enemy, the enemy of my enemy is my friend}~\cite{Schwartz:2010}, creates social tensions which social actors, according to balance theory, tend to resolve. Whereas structural balance is inherent to many real-world networks~\cite{Facchetti:2011,EasleyKleinberg,ZhengZengWang:2015}, classical balance theory does not explain \emph{how} the system reaches the balanced state. Dynamics of weights $a_{ij}$ that lead to their structural balance are beyond the scope of this tutorial
and can be found e.g. in~\cite{Hummon:2003,Antal:2005,Antal:2006,Kulakowski:2007,Abell:2009,Marvel:2011,FriedkinJiaBullo:2016-1,Parravano:16}.

It can be seen that the nodes of a structurally balanced sign-symmetric complete graph split into two factions, or ``hostile camps''~\cite{ProMatvCao:2016} as follows. Consider an arbitrary agent $i$ and let the set $V_1$ consist of $i$ and his/her friends $V_1=\{i\}\cup\{j\ne i:a_{ij}>0\}$, while the set $V_2$ consists of his/her enemies $V_2=\{j\ne i:a_{ij}<0\}$. By definition of the structural balance, every two friends, as well as every two enemies, of agent $i$ are friends, i.e. $a_{jk}>0$ whenever $(j,k)\in (V_1\times V_1)\cup (V_2\times V_2)$ and $j\ne k$. Similarly, a friend of $i$ and an enemy of $i$
are always enemies: $a_{jk}<0$ when $(j,k)\in (V_1\times V_2)\cup (V_2\times V_1)$ and $j\ne k$. The converse statement is also valid: if the set of nodes $V$ can be decomposed into two disjoint sets $V_1,V_2$ with the aforementioned properties, then the graph is structurally balanced. Indeed, in each triad two nodes belong to the same ``camp''.
The remaining node can either belong to the same camp or the opposite camp, which corresponds, respectively, to the triads of types (a) and (b) in Fig.~\ref{fig.triad}.
This result, established in~\cite{Harary:1953} and called the ``balance theorem'',
inspires the following general definition.

\begin{defn}(Structural balance)\label{def.balance}
 A general signed graph is $G=(V,E,A)$ is \emph{structurally balanced} if the set of its nodes can be divided into such disjoints subsets $V=V_1\cup V_2$ that
for any pair $i,j\in V$, $i\ne j$,
 \[
 \begin{cases}
 a_{ij}\ge 0,\quad \text{if}\quad (i,j)\in (V_1\times V_1)\bigcup (V_2\times V_2),\\
 a_{ij}\le 0,\quad \text{if}\quad (i,j)\in (V_1\times V_2)\bigcup (V_2\times V_1).
 \end{cases}
 \]
 \end{defn}

Note that one of the ``camps'' $V_i$ may be empty (this holds in the case of a usual weighted graph, where the matrix $A$ is nonnegative). Unlike the case of a complete graph, balance in each triad is insufficient for structural balance in the graph. To formulate a necessary and sufficient condition, we introduce two auxiliary concepts.
\begin{defn}
 For a sign-symmetric graph $\g=(V,E,A)$, consider an undirected graph $\hat \g=(V,\hat E,A+A^{\top})$ obtained from $\g$ by ``mirroring'' each directed arc.\footnote{Since $a_{ij}a_{ji}\ge 0$ for $i\ne j$, $(A+A^{\top})_{ij}\ne 0$ if and only if $|a_{ij}|+|a_{ji}|\ne 0$. In other words, $(i,j)\in\hat E\Leftrightarrow (j,i)\in\hat E\Leftrightarrow (i,j)\in E\lor (j,i)\in E$, so the undirected graph $\hat \g$ contains all directed arcs $(i,j)$ of $\g$ together with their ``mirrors'' $(j,i)$.} A \emph{semiwalk} $v_0,\ldots,v_n$ in the graph $\g$ is a sequence of nodes, corresponding to a walk in $\hat \g$; if $v_0=v_n$, this semiwalk is called \emph{semicycle}.
\end{defn}
\begin{defn}
We call a walk $v_0,\ldots,v_n$ is a signed graph $\g$ \emph{positive} (respectively, \emph{negative}) if the product of all weights on the connecting arcs $a_{v_0v_1}a_{v_1v_2}\ldots a_{v_{n-1}v_n}$ is positive (negative). The sign of a semiwalk in a sign-symmetric graph $\g$ is its sign as a walk in the corresponding undirected graph $\hat\g$.
\end{defn}

\begin{figure}[h]
\center
\includegraphics[width=0.33\columnwidth]{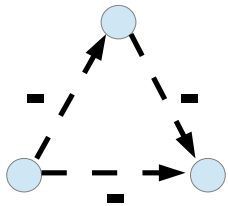}
\caption{A graph without cycles, no structural balance}
\label{fig.triad+}
\end{figure}

It can be shown (using e.g. induction on the walk's length) that a negative walk in a structurally balanced graph connects the nodes from different camps, whereas a positive walk starts and ends at the same camp. In particular, for structural balance it is \emph{necessary} that the all cycles are positive (for cycles of length 2, this implies the sign symmetry of the graph).
This is condition is also \emph{sufficient} for strongly connected graphs~\cite{Altafini:2013}, however, without strong connectivity, even graphs with $n=3$ nodes can be structurally imbalanced and have no cycles (see Fig.~\ref{fig.triad+}).
By noticing that the undirected graph $\hat\g$, corresponding to a structurally balanced graph $\g$, is also structurally balanced, one shows that in fact all \emph{semicycles} have to be positive,
and this condition appears to be sufficient for structural balance~\cite{CartwrightHarary:1956}.
\begin{lem}\footnote{The necessary and sufficient condition from this lemma is often used as a definition of structural balance~\cite{CartwrightHarary:1956}.}
If a signed graph is structurally balanced, then all its cycles are positive (in particular, the graph is sign-symmetric). This condition is also \emph{sufficient} for structural balance when the graph is strongly connected. In general, for structural balance it is necessary and sufficient that the graph is sign-symmetric and all its \emph{semicycles} are positive.
\end{lem}

The concept of structural balance has been extended in various directions, e.g. the more general condition of \emph{weak balance} allows clustering into more than two factions~\citep{Davis:1967}, where members of each factions are friends and the members of different factions are enemies. For undirected graphs, the weak balance is equivalent to the absence of cycles with exactly one negative arc~\cite{Davis:1967}. A natural question, concerned with imbalanced signed graph,
is how to measure the ``level of imbalance'' (or how close the graph is to balance). Different measures of imbalance have been proposed in~\cite{Zaslavsky:1987,ZhengZengWang:2015,Facchetti:2011,EasleyKleinberg}.

Following \cite{Altafini:2013}, we define the \emph{Laplacian} matrix $L=L[A]$ of the signed graph $\g=(V,E,A)$ as follows
\be\label{eq.laplacian-sign}
L[A]=(l_{jk}),\; l_{jk}=\begin{cases}
-a_{jk},\,j\ne k\\
\sum_{m=1}^N|a_{jm}|, j=k.
\end{cases}
\ee

In the case where $A$ is nonnegative,~\eqref{eq.laplacian-sign} coincides with the conventional Laplacian matrix of a weighted graph~\cite{ProTempo:2017-1}. As implied by the Gershgorin Disc Theorem, $L[A]$ has no
eigenvalues in the closed left half-plane
$\bar\cc_-=\{\lambda\in\cc:\re\lambda\le 0\}$ except for, possibly, $\lambda=0$.
Unlike the unsigned case, in general $L[A]$ can have no zero
eigenvalue, in which case the matrix $(-L[A])$ is Hurwitz.

The following elegant result~\cite{Altafini:2013} establishes a relation between the Laplacian and structural balance.
\begin{lem}\label{lem.altaf1}
For a strongly connected signed graph $\g=(V,E,A)$, $\la=0$ is an eigenvalue of $L[A]$ if and only if
the graph $\g$ is structurally balanced.
\end{lem}

As will be shown, the sufficiency part in Lemma~\ref{lem.altaf1} in fact does not rely on the strong connectivity. The ``gauge transformation''~\cite{Altafini:2013} introduced in the next subsection
transforms the Laplacian of a structurally balanced graph $L[A]$ into the matrix $L[A^{|\cdot|}]$, where $A^{|\cdot|}=(|a_{ij}|)$. In other words, $L[A]$ is similar to the Laplacian of an (unsigned) weighted graph, which always has an eigenvalue at $0$.

\subsection{Altafini's model of opinion formation}

Balance theory suggests that a natural outcome of the opinion formation process should somehow reflect the partition of the network into two opposing factions~\citep{Altafini:2012}. In the models offered in~\cite{Altafini:2012,Altafini:2013}, structural balance of a signed network leads to \emph{polarization} (``bipartite consensus'') of the opinions:
opinions in each faction reach consensus, and their consensus values are equal in modulus yet differ in sign.

The original\footnote{In this tutorial, we confine ourselves to the linear Altafini model and do not consider its nonlinear counterparts~\cite{Altafini:2012,Altafini:2013,ProMatvCao:2016,ProMatvCao:2014,ProCao:2014}, whose properties are in fact very similar.} Altafini model~\cite{Altafini:2013} is coincident with the Abelson model~\eqref{eq.abelson1},
however the matrix $A(t)$ need not be nonnegative, and the corresponding Laplacian is understood in the sense of~\eqref{eq.laplacian-sign}
\be\label{eq.altaf}
\begin{gathered}
\begin{aligned}
\dot x_i(t)=\sum_{j=1}^n|a_{ij}(t)|(x_j(t)\sgn a_{ij}(t)-x_i(t))=\\
=\quad \sum_{j=1}^n(a_{ij}(t)x_j(t)-|a_{ij}(t)|x_i(t))\,\forall
i,\quad \overset{\eqref{eq.laplacian-sign}}{\Longleftrightarrow}
\end{aligned}\\
\Longleftrightarrow\dot x(t)=-L[A(t)]x(t),\quad t\ge 0.\\
\end{gathered}
\ee

To ensure the existence and uniqueness of the solution for any $x(0)$, it suffices to assume that $A(\cdot)$ is locally $L_1$-function.
Similar to the Abelson model, solutions of the Altafini model are globally bounded; it can be shown~\cite{ProMatvCao:2016} that $M(t)=\max_i|x_i(t)|$ is a non-increasing function
and hence $|x_i(t)|\le M(0)$ for any $i$ and $t\ge 0$, entailing the following proposition.
\begin{prop}\label{prop.converge-alt0}
System~\eqref{eq.altaf} is Lyapunov stable.
\end{prop}

As we will see, unlike the unsigned case the system~\eqref{eq.altaf} can, in general, be \emph{asymptotically stable}.

\subsubsection{The case of a time-invariant signed graph}

We start with analysis of the static Altafini model ($A(t)\equiv A$ is a constant matrix). Similar to the static Abelson model, the opinions always converge.
\begin{prop}\label{prop.converge-alt}
For any matrix $A$ and initial condition $x(0)$, there exist finite limits $\bar x_i=\lim\limits_{t\to\infty}x_i(t)$.
\end{prop}

Proposition~\ref{prop.converge-alt} is immediate from Proposition~\ref{prop.converge-alt0}. If $L[A]$ has an eigenvalue
at $0$, the Jordan cells corresponding to it are trivial in view of the Lyapunov stability. Since all other eigenvalues of $(-L[A])$ are stable, this implies the existence of $\lim\limits_{t\to\infty}e^{-L[A]t}$. $\square$

Proposition~\ref{prop.converge-alt} can be also proved by using the \emph{lifting approach}~\cite{Hendrickx:14}, showing that the Altafini model is equivalent to Abelson's model with $2n$
agents $1,\ldots,n$ and $1',\ldots,n'$, where agent $i'$ is the antipode of agent $i$ and has opinion $x_{i'}=(-x_i)$. Some other results, concerned with the behavior of Altafini's model, can also be derived by using this lifting technique.  Note that the arguments, used in~\cite{ProTempo:2017-1} to prove the convergence of static Abelson models, cannot be
extended to the signed case since $L[A]$ is no longer a $M$-matrix.

The Altafini model has the most interesting behavior when the graph $\g=(V,E,A)$ is \emph{structurally balanced}. Consider the decomposition of $V$ into two ``hostile camps'' $V=V_1\cup V_2$
and the following mapping, referred in~\cite{Altafini:2013} to as the \emph{gauge transformation}
\be\label{eq.gauge}
x_i\mapsto y_i=\delta_ix_i,\quad \delta_i=
\begin{cases}
+1,\; &i\in V_1\\
-1,\; &i\in V_2.
\end{cases}
\ee
By noticing that $\delta_i\delta_ja_{ij}=|a_{ij}|$,~\eqref{eq.altaf} shapes into
\be\label{eq.altaf-gauge}
\dot y_i(t)=\sum_{j=1}^n|a_{ij}|(y_j(t)-y_i(t))\quad\forall i,
\ee
that is, the vector $y(t)$ obeys the Abelson model with the (nonnegative) weight matrix $A^{|\cdot|}=(|a_{ij}|)$ and
\[
L[A]=\Delta L\left[A^{|\cdot|}\right]\Delta,\quad \Delta=\Delta^{-1}=\diag(\delta_1,\ldots,\delta_n).
\]
In particular, if the graph $\g$ has a directed spanning tree~\cite{ProTempo:2017-1}, then consensus is established in the system~\eqref{eq.altaf-gauge} $y_i\xrightarrow[t\to\infty]{}p^{\top}y(0)$,
where $p\ge 0$ is the only left eigenvector of $L[A^{|\cdot|}]$ such that $p^{\top}L[A^{|\cdot|}]=0$ and $p^{\top}\mathbbm{1}_n=1$ (the vector of ``social powers''~\cite{ProTempo:2017-1}).
This implies the following polarization property of the solutions, referred also to as the \emph{bipartite consensus}~\cite{Altafini:2013}
\be\label{eq.bipart}
\lim_{t\to\infty} x_i(t)=
    \begin{cases}
    w^{\top}x(0),\quad &i\in V_1\\
    -w^{\top}x(0),\quad &i\in V_2,
    \end{cases}
\ee
where $w=\Delta p$ is some non-zero vector (not necessarily positive). In other words, for almost all initial conditions opinions in the ``hostile camps'' reach consensus at the opposite values.

If $\g$ is structurally balanced but has no directed spanning tree, the opinions in~\eqref{eq.altaf-gauge} (and thus also in~\eqref{eq.altaf}) split into several clusters, determined
by the graph's matrix of ``spanning forests''~\cite{AgaevChe:2014}.

\begin{figure}
\centering
\includegraphics[width=0.33\columnwidth]{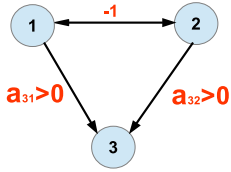}
\caption{A structurally imbalanced quasi-strongly (yet not strongly) connected graph}
\label{fig.1}
\end{figure}

Consider now the case of structurally \emph{imbalanced} graph. In the case where $\g$ is \emph{strongly connected}, Lemma~\ref{lem.altaf1} implies that the system~\eqref{eq.altaf} is exponentially stable since $(-L[A])$ is Hurwitz. In this case, opinions reach consensus at zero value independent of the initial condition.
This statement is \emph{not valid} for graphs with multiple strongly component component, as illustrated by the simple example with $n=3$ agents~\cite{ProMatvCao:2016}.
Let $a_{12}=a_{21}=-1$ and $a_{31},a_{32}>0$, so the graph $\g$ has a directed spanning tree and is structurally imbalanced (Fig.~\ref{fig.1}).
The Altafini system
\ben
\begin{split}
\dot x_1=(-x_2-x_1),\,\dot x_2=(-x_1-x_2),\\
\dot x_3=a_{31}x_1+a_{32}x_2-(a_{31}+a_{32})x_3,
\end{split}
\een
has equilibria $(\xi,-\xi,\rho\xi)$, with $\rho=(a_{31}-a_{32})/(a_{31}+a_{32})\in (-1;1)$, $\xi\in\r$, exhibiting thus neither polarization nor stability. Such a behavior is a special case of ``interval bipartite consensus\footnote{The work~\cite{Meng:16-2} deals with a quasi-strongly connected graphs with multiple root nodes (like nodes 1 and 2 in Fig.~\ref{fig.1}), constituting
a closed strongly connected component~\cite{ProTempo:2017-1} and reaching consensus in modulus (see Remark~\ref{rem.mod-conse} below). Denoting their steady modulus value by $\theta=\theta(x(0))\ge 0$, rhe ``interval bipartite consensus'' is defined in~\cite{Meng:16-2} as the convergence of all opinions $x_i(t)$ as $t\to\infty$ to the interval $[-\theta,\theta]$.}''~\cite{Meng:16-2}. We do not consider the structure of the clusters, arising in~\eqref{eq.altaf} in the general case of imbalanced yet not strongly connected graphs
and conditions, ensuring stability in this case, for further reading see~\cite{Meng:16-2,ProMatvCao:2016}.

The results concerning the behavior of static Altafini's model are summarized in the following.
\begin{thm}
For the Altafini model over a static graph $\g=(V,E,A)$, the following statements hold:
\begin{enumerate}[(a)]
\item if the graph has an directed spanning tree and is structurally balanced, the opinions polarize~\eqref{eq.bipart};
\item if the graph is \emph{strongly connected} and is structurally imbalanced, the system is stable;
\item in general, the opinions converge and can split into several clusters, whose number and structure depend on the graph $\g$.
\end{enumerate}
\end{thm}

\begin{rem}\label{rem.mod-conse}
In the cases (a) and (b) opinions reach \emph{consensus in modulus} in the sense that
\be\label{eq.mod-conse}
\lim_{t\to\infty} |x_1(t)|=\ldots=\lim_{t\to\infty} |x_n(t)|.
\ee
\end{rem}

\subsubsection{The dynamic graph case}

The case of a general time-varying matrix $A(t)$ has not been fully studied, which is not surprising since necessary and sufficient conditions for convergence and consensus are still elusive even for the time-varying Abelson model ($a_{ij}\ge 0$). Obviously, if the graph $\g[A(t)]$ is structurally balanced and the decomposition $V=V_1\cup V_2$ is \emph{time-invariant} (that is, the friendship and enmity relations between each two agents are constant), the model reduces to Abelson model~\eqref{eq.altaf-gauge} via the gauge transformation~\eqref{eq.gauge}. In general, such a transformation is not possible, nevertheless the properties of the Altafini model~\eqref{eq.altaf} and the associated Abelson model~\eqref{eq.altaf-gauge} are closely related.

The following counterpart of Lemma~\ref{lem.converge2} has been established in~\cite{Hendrickx:14,ProMatvCao:2016,ProCao:2017}.
\begin{lem}\label{lem.converge2+}
Suppose that the gains $a_{ij}(t)$ satisfy the following type-symmetry condition
\be\label{eq.symmetry}
K^{-1}|a_{ji}(t)|\le |a_{ij}(t)|\le K|a_{ji}(t)|\quad \forall t\ge 0,
\ee
where $K\ge 1$ is a constant. Then the limit $\xi_i=\lim_{k\to\infty} |x_i(k)|$ exists for each $i$. If agents $i$ and $j$ \emph{interact persistently} in the sense that
\[
\int_{0}^{\infty} |a_{ij}(t)|dt=\infty,
\]
then the moduli of their final opinions coincide $\bar \xi_i=\bar\xi_j$. If the graph of persistent interactions is connected, consensus in modulus~\eqref{eq.mod-conse} is established.
\end{lem}

Similar to Lemma~\ref{lem.converge2}, Lemma~\ref{lem.converge2+} can be extended to graphs satisfying a more general cut-balance condition~\cite{ProCao:2017}.
Lemma~\ref{lem.converge2+} implies, in particular, that if~\eqref{eq.symmetry} holds consensus in~\eqref{eq.altaf-gauge} is equivalent to consensus in modulus in~\eqref{eq.altaf}.
This consensus in modulus can be either polarization~\eqref{eq.bipart} (with some $w\ne 0$) or asymptotic stability; necessity and sufficient conditions for both types of behavior
has been given in~\cite{Hendrickx:14,ProMatvCao:2016}. It is remarkable that Lemma~\ref{lem.consensus2} cannot be extended to Altafini's model in the same way: the UQSC condition for the matrix
$A^{|\cdot|}$ implies consensus in~\eqref{eq.altaf-gauge}, however, in general it implies neither consensus of the opinions' absolute values~\eqref{eq.mod-conse}, nor even their convergence~\cite{ProMatvCao:2016}. Consensus in modulus is provided by the uniform \emph{strong} connectivity.
\begin{lem}\label{lem.consensus2+}
Let $A(\cdot)$ be bounded $|a_{ij}(t)|\le M$ and there exist $\ve,T>0$ such that the following graphs
\be\label{eq.graph-union2}
\g_{\ve,T}(t)=\g\left[\int_t^{t+T}A^{|\cdot|}(s)ds\right],\quad t\ge 0
\ee
are \emph{strongly} connected for any $t\ge 0$. Then consensus in modulus~\eqref{eq.mod-conse} is established.
\end{lem}

Unlike the type-symmetric case from Lemma~\ref{lem.converge2+}, in the situation of Lemma~\ref{lem.consensus2+} it is not easy to give conditions for polarization and stability.
To the best of the authors' knowledge, this problem has been solved only under additional assumptions~\cite{Hendrickx:14,LiuChenBasar:2017,MengMengHong:2018}.

\subsection{Extensions and related works}

In~\cite{AltafiniLini:15}, a generalization of the Altafini models has been studied as follows
\[
\dot x_i(t)=-\sigma_i x_i(t)+\sum_{j=1}^n a_{ij}x_j(t),
\]
where $\sigma_i>0$ are constant ``degradation rates'', $a_{ii}=0$ and $a_{ij}$ can be both positive and negative for $i\ne j$. The main concern is the reaching of \emph{sign consensus}, or ``unanimity''~\cite{AltafiniLini:15} of opinions, that is, their convergence to the cone obtained by union of the positive and the negative orthant $\r_+^n\cup\r_-^n$.
The key property, providing such a relaxed consensus, is \emph{eventual nonnegativity} of the matrix $A$ (a matrix is eventually nonnegative if all its powers $A^k$ are non-zero
and at least one of them is nonnegative). In~\cite{CeragioliAltafini:2016}, an interesting model has been proposed that combines the ideas of antagonistic interactions and bounded confidence
(a smooth counterpart of this model has been also studied in~\cite{ProCao16-3}). Extending the idea of gauge transformation, distributed continuous-time protocols sorting a given list of real numbers have been proposed in~\cite{Altafini:14-sort}.

Along with continuous-time model~\eqref{eq.altaf}, the ``discrete-time Altafini model'' can be considered
\be\label{eq.altaf-d}
\begin{gathered}
x(k+1)=W(k)x(k)\in\r^n,
\end{gathered}
\ee
where the matrix $W^{|\cdot|}=(|w_{ij}(k)|)$ is stochastic and $w_{ii}(k)\ge 0$, while $w_{ij}(k)$ for $i\ne j$ can be negative. Properties of the discrete-time model~\eqref{eq.altaf-d} are similar to the properties of its continuous-time counterpart; e.g. consensus in modulus can be proved under the assumption of repeated \emph{strong} connectivity.
For this reason we do not consider the relevant theory and refer the reader to recent works~\cite{XiaCaoJohansson:16,MengShiCao:16,LiuChenBasar:2017,ProCao2017-3}.

Among other extensions, gossip-based models with antagonistic interactions~\cite{ShiJohanssonJohansson:13,ShiBarasJohansson:15,ShiBarasJohansson:16} should be mentioned. Unlike Altafini's model,
some of these models~\cite{ShiBarasJohansson:15,ShiBarasJohansson:16} provide polarization under the assumption of \emph{weak} structural balance~\cite{Davis:1967}.

%% file: 6conclus.tex
\section{Conclusions and Future Works}\label{sec.conclus}

The models describing social processes are numerous. It will not be
an exaggeration to say that almost every week a novel model appears.
When this tutorial was started, many of the papers referred in it
had not been even written. Confining ourselves to a special class of
dynamic models, we clearly realize that even this class remain
partially uncovered by this tutorial. For instance, we do not
consider models with \emph{quantized} communication among the agents
(that is, information an agent displays to the others is limited to
a finite set of
symbols)~\cite{Urbig:2003,Martins:13,CeragioliFrasca:18}. Focusing
on stability and convergence problems, we avoid other important
properties of opinion formation models such as e.g. their
controllability~\cite{Dietrich:18,Masuda:2015} and
identifiability~\cite{Scaglione:16,RavazziTempoDabbene:2018}.
Processes closely related to opinion formation, e.g. the dynamics of
reflected
appraisals~\cite{Bullo:2013,FriedkinJiaBullo:2016,ChenLiuBelabbasXuBasar:17,YeBasarAnderson:17},
have been also excluded from the consideration. A very recent
direction of research, opened by~\cite{HendrickxMartin:17}, deals
with \emph{open} models of multi-agent systems, which can be joined
and left by agents. This approach opens up the perspective of
modeling real social media, where agents interact via web forums,
chats and blogs, and other \emph{temporal} social networks that can
loose and acquire not only connections among the nodes, but also
nodes themselves. To make social networks more resilient against
malicious attacks, it is important to understand mechanisms of
``misbehavior'' such as e.g. rumors, fake news and misinformation
spreading~\cite{AcemogluOzdaglar:2011,DelVicario:16}. Models
proposed to describe \emph{belief systems}~\cite{Converse:1964} and
portraying the evolution of opinions on multiple interrelated
topics~\cite{FriedkinPro2016,Parsegov2017TAC,XiongLiuWangWang:17,NoorazarArxiv:16}
also remain uncovered.

To overview all cutting-edge models in a journal paper is hardly possible, and without any doubts, sooner or later textbooks and monographs on social dynamics modeling will be published.
The goal of this tutorial is to unveil the new field, lying in the frontier between systems theory and social science, to the broadest audience, and we hope that it will be helpful
for researchers, starting working in this area.


In spite of recent progress in mathematical studies of opinion formation models, we still know very little about real processes in society. All of the existing models describe only one of the numerous facets of the social life, and most of focus on one special effect or property of social interactions (stubborness, homophily, xenophobia etc.)
The ancient fable about blind men touching an elephant teaches us that this reductionist approach sometimes fails to
give a real picture of the real phenomenon. Should we continue studying simplified models or seek for a complex holistic models of social influence?
Which of the existing ``simple'' models is closer to reality? Should we use different models in different situations?

Evidently, mathematics cannot answer these questions; the only way to get realistic models is to work with data.
Such data can be collected during social experiments. Nowadays, a huge amount of data is available in online social media.
Although we witness some examples of successful interdisciplinary collaborations in experimental validation of dynamic models~\cite{FriedkinBullo:17,Kerckhove:16},
the current level of collaboration among between the communities systems and control theorists, sociologists and data scientists is insufficient.
From our viewpoint, the toughest challenge is to unite the forces of these communities, elaborating a common language and instruments to be used in qualitative
and quantitative studies of social dynamics.

\section*{Afterword by Anton V. Proskurnikov}

This paper was conceived by Dr. Roberto Tempo and myself in 2016 as
a survey, giving an overview of social dynamics models, scattered in
mathematical, physical, sociological and engineering literature,
from the systems and control viewpoint. Soon we realized that such a
survey will be appreciated only by researchers, working on consensus
and coordination of multi-agent networks, whereas our purpose was to
open the exciting field of ``social systems theory'' to the broadest
audience. It was decided to transform the survey into a tutorial,
starting from the very beginning and introducing all necessary
mathematical concepts. We have followed this plan in Part I, dealing
with classical models of opinion formation.

Roberto's tragic and abrupt decease in January 2017 has made me postpone the submission of Part II for more than half a year and seriously restructure the text. Some sections, promised in Part I (e.g. distributed algorithms for network analysis and detailed discussion of emerging directions) have been discarded or shortened.
Besides this, I have decided to include some very recent results, trying to keep a reasonable balance between tutorial and survey functions of this paper.
Since the relevant mathematical techniques are too complicated to be considered, only some of the results are accompanied by full proofs.

Whether we like this or not, this tutorial is written by control theorists and primarily deals with problems, addressed by the systems and control community. I have received several emails, asking
about the relation between the ``strange'' problems we consider and ``actual'' theory of social networks. Although I have done my best to explain the difference between the problems, models and approaches considered in this tutorial, and those studied by other communities (e.g. social network analysts, data scientists and statistical physicists), I have to repeat the key theses. We consider not \emph{social networks} themselves, but rather \emph{dynamical processes} over them, namely, dynamics of \emph{opinion formation}. Among the dynamic models,
we focus on agent-based models with continuous opinions, whereas statistical and discrete opinion models remain uncovered by this paper.

I am grateful to all colleagues, encouraging me to finish this important work, being one of the last Roberto's projects. I was especially encouraged by receiving messages from young researchers, urging me with impatience to publish its draft. I am indebted to Francoise Lamnabhi-Lagarrigue, Editor in Chief of Annual Reviews in Control, for inviting us to contribute this tutorial, and to my colleagues Noah Friedkin, Andreas Fl\"ache, Francesco Bullo, Paolo Frasca, Claudio Altafini, Julien Hendrickx, Fabrizio Dabbene and Chiara Ravazzi for fruitful discussions.